\documentclass[12pt]{article}

\usepackage{graphicx}
\begin{document}

\pagestyle{empty}
\vspace{-2in}
\begin{flushright}
RM3-TH/02-22 \\
ROMA-1350/02
\end{flushright}
\vskip 1cm \centerline{\Large{\bf \boldmath$b\to s$ Transitions: A
    New Frontier}}\vskip0.2cm
  \centerline{\Large{\bf for Indirect SUSY Searches}} \vskip 1.4cm
\centerline{{\bf M.~Ciuchini$^{1}$, E.~Franco$^2$, A.~Masiero$^3$ and
    L.~Silvestrini$^2$}} \vskip 0.4cm \centerline{$^1$ {\sl INFN
    Sezione di Roma III and Dip. di Fisica, Univ. di Roma Tre,}}
\centerline{{\sl Via della Vasca Navale 84, I-00146 Rome, Italy.}}
\centerline{$^2$ {\sl INFN Sezione di Roma and Dip. di Fisica, Univ.
    di Roma ``La Sapienza'',}} \centerline{{\sl P.le A. Moro 2,
    I-00185 Rome, Italy.}}  \centerline{$^3$ {\sl Dip. di Fisica ``G.
    Galilei'', Univ. di Padova and INFN,}} \centerline{{\sl Sezione di
    Padova, Via Marzolo 8, I-35121 Padua, Italy.}}  \vskip 1cm
\abstract{The present unitarity triangle fit, whose essential input is
  represented by the $s \to d$ and $b \to d$ transition processes,
  fully agrees with the SM. However, most of the phenomena involving
  $b \to s$ transitions are still largely unexplored and hence $b \to
  s$ phenomenology still constitutes a place for new physics
  manifestations, in spite of the tremendous experimental and
  theoretical progress on $B \to X_s \gamma$. We perform a systematic
  study of the CP conserving and violating SUSY contributions to $b
  \to s$ processes in a generic MSSM.  We consider gluino exchange
  contributions including NLO QCD corrections and lattice hadronic
  matrix elements for $\Delta B = 2$ and $\Delta B = 1$ processes. We
  take into account all available experimental information on
  processes involving $b \to s$ transitions ($B \to X_s \gamma$, $B
  \to X_s \ell^+ \ell^-$ and the lower bound on the $B_s - \bar B_s$
  mass difference $\Delta M_s$). We study the correlations among the
  relevant observables under scrutiny at present or in a not too far
  future: $\Delta M_s$ and the amount of CP violation in $B \to \phi
  K_s$, $B_s \to J/\psi \phi$, $B \to X_s \gamma$. In particular we
  discuss the recent data by BaBar and BELLE on the time-dependent CP
  asymmetry in the decay $B \to \phi K_s$ which suggest a deviation
  from the SM expectation. Our results show that the processes
  involving $b \to s$ transitions represent a splendid opportunity to
  constrain different MSSM realizations, and, even more important,
  that they offer concrete prospects to exhibit SUSY signals at $B$
  factories and hadron colliders in spite of all the past frustration
  in FCNC searches of new physics hints.}
\newpage
\pagestyle{plain}
\setcounter{page}{1}

\section{Introduction}
The Standard Model (SM) of electroweak and strong interactions is
extremely successful in describing all available data on particle
physics up to the highest energies reached presently at colliders.
However, its least understood aspects concern electroweak and flavour
symmetry breaking, i.e.~the Higgs sector and fermion masses and
mixings. Supersymmetry (SUSY) is the only known low-energy extension
of the SM which, in addition to stabilizing the Higgs sector against
radiative corrections, allows for gauge coupling unification at high
energies, without spoiling the agreement with precision electroweak
data. However, on its own it does not add any understanding of flavour
symmetry breaking. Indeed, flavour physics is a very stringent test of
SUSY extensions of the SM: in its general form, the Minimal
Supersymmetric Standard Model (MSSM) contains more than 100 new
parameters in the flavour sector, which can cause Flavour Changing
Neutral Current (FCNC) and CP violating processes to arise at a rate
much higher than what is experimentally
observed~\cite{antichi}--\cite{Bagger:1997gg}.

Within the SM, the only sources of flavour and CP violation in the
hadronic sector arise from the Cabibbo-Kobayashi-Maskawa (CKM) quark
mixing matrix. Recent progresses in experimental results and in
theoretical methods allow for a very successful determination of the
CKM parameters via the so-called Unitarity Triangle (UT) fit, which
combines all presently available informations on flavour and CP
violation. The success of the SM UT fit is a clear signal that present
experimental data do not favour new generic sources of FCNC and CP
violation. However, we think that it is premature to draw the
conclusion that the room available for New Physics (NP) in FCNC and CP
violating processes has shrunk to the point that no significant
departure from SM expectations is foreseable within such phenomena.

A closer look at the UT fit reveals that NP
contributions to $s \to d$ and $b \to d$ transitions are strongly
constrained, while new contibutions to $b \to s$ transitions do not
affect the fit at all, unless they interfere destructively with the SM
amplitude for $B_s - \bar B_s$ mixing, bringing it below the present
lower bound of $\sim 14$ ps$^{-1}$.  It is certainly true that other
processes not directly involved in the UT fit, in particular $B \to X_s
\gamma$ and, to a lesser extent, $B \to K \pi$, $B \to \phi K_s$, $B \to
X_s \ell^+ \ell^-$ and $B_s \to \ell^+ \ell^-$ decays represent a
powerful constraint on any NP in $b \to s$ transitions. However, the
celebrated $B \to X_s \gamma$ decay mostly constrains the helicity
flipping contributions to the $b \to s$ transition and, as we shall
see in the following, in the case of SUSY the effect of these
constraints is not as dramatic as it is for $s \to d$ and $b \to d$
transitions, and plenty of room is left for SUSY contributions to
interesting observables in this sector.

In this work we intend to make a systematic study of the SUSY
contributions to the (CP conserving and violating) $b \to s$
transitions in the context of a {\it generic} MSSM model with R
parity.  We see three main motivations to pursue yet another analysis
on the thorougly explored issue of FCNC and MSSM.

i) Thanks to the advent of the new B factories, our experimental 
knowledge
concerning the above processes has greatly improved in this last period
and, in addition, we expect new results on the $B_s - \bar B_s$ mixing
soon at Run II of the Tevatron from CDF and later on BTeV and LHCB. 
Hence,
a more thorough investigation of the $b \to s$ transition both in 
$\Delta B=2$ and $\Delta B=1$ processes is now mandatory.

ii) While for $\Delta B=2$ processes a refined treatment of the gluino
exchange contributions including the NLO QCD evolution for the 
four-quark
operators and a lattice computation of the hadronic matrix elements is 
now
available~\cite{Becirevic:2001jj}, the same does not apply to the 
$\Delta B=1$
case. Moreover correlations among different observables in $b \to s$
physics have still to be largely explored in the MSSM context. 

iii) Finally, it was recently pointed
out~\cite{Moroi:2000tk,Chang:2002mq} that the large mixing(s) in the
neutrino sector may imply the presence of large mixing angles in the
right-handed down-type quarks in GUT's where the latter are unified
with the lepton doublets. Although we cannot witness the presence of
such large mixings in our experiments given that they are present in
the right-handed charged hadronic currents, 
we can still have a visible implication in
low-energy physics because of the possible large mixing between
right-handed bottom and strange squarks through radiative corrections
induced by the large top Yukawa coupling.

Although this latter motivation iii) would push us towards the choice
of particular patterns of squark mass matrices with only the $\tilde
b_R - \tilde s_R$ entry largely enhanced (see
ref.~\cite{Harnik:2002vs}), in view also of the previous two
motivations we think it more interesting not to commit ourselves to
any specific choice of the squark masses, but rather to keep our
analysis in the MSSM as general as possible. Obviously, then the
enhanced $\tilde b_R - \tilde s_R$ case turns out to be just a
particular case and we are able to appreciate even more its
characterizing features when we embed it in a {\it generic} low-energy
SUSY extension of the SM. Needless to say, there is a price to pay to
allow ourselves to be so general: the complete ignorance of the squark
mass textures prevents us from using the basis of the squark physical
(mass) eigenstates, but rather we have to make use of an efficient
parametrization of the generic squark mass matrix diagonalization.
Such a tool has been available for a long time \cite {hall}: we
parametrize the new sources of flavour and CP violation in the
hadronic sector present in a generic MSSM choosing the so-called
Super-CKM basis. In this basis, all gauge couplings involving SUSY
partners of quarks have the same flavour structure as the
corresponding quark couplings. FCNC and CP violation arise then from
off-diagonal terms in squark mass matrices. These are conveniently
expressed as $(\delta_{ij})_{AB}\equiv (\Delta_{ij})_{AB}/m^2_{\tilde
  q}$, where $(\Delta_{ij})_{AB}$ is the mass term connecting squarks
of flavour $i$ and $j$ and ``helicities'' $A$ and $B$, and
$m_{\tilde q}$ is the average squark mass. In the absence of any
horizontal symmetry and for a generic SUSY breaking mechanism, one
expects $(\delta^d_{ij})_{LL} \leq \mathcal{O} (1)$,
$(\delta^d_{ij})_{RR} \leq \mathcal{O} (1)$, $(\delta^d_{ij})_{LR}
\leq \mathcal{O} (m_{d_k}/m_{\tilde q})$ and $(\delta^d_{ij})_{RL}
\leq \mathcal{O} (m_{d_k}/m_{\tilde q})$, with $k=$max$(i,j)$. The
last two inequalities are also imposed by the requirement of avoiding
charge and colour breaking minima as well as unbounded from below
directions in scalar potentials~\cite{Casas:1996de}. We argued above
that the UT fit poses stringent constraints on NP contributions to $s
\to d$ and $b \to d$ transitions, which in the MSSM are governed by
$(\delta^d_{12})_{AB}$ and $(\delta^d_{13})_{AB}$ respectively.
Indeed, detailed analyses carried out in SUSY at a level of accuracy
comparable to the SM UT fit (NLO QCD corrections, Lattice QCD hadronic
matrix elements) have shown that one must have $(\delta^d_{12})_{AB}$
and $(\delta^d_{13})_{AB}$ much smaller than what naively
expected~\cite{Becirevic:2001jj,Ciuchini:1998ix}. It is therefore
reasonable to assume that, either due to the effect of some horizontal
symmetry or to the explicit form of SUSY breaking,
$(\delta^d_{12})_{AB}\sim (\delta^d_{13})_{AB}\sim 0$. The purpose of
this paper is to thoroughly analyze, once again at the same level of
accuracy of SM investigations, constraints on $(\delta^d_{23})_{AB}$
from available data and possible effects in present and future
measurements. As we shall see, in this case $(\delta^d_{23})_{AB}$ at
the level of what naively expected are certainly allowed by present
data. Moreover, if some recent experimental results, such as the CP
asymmetry in $B \to \phi K_s$ decays~\cite{Aubert:2002nx,Stocchi:2002yi}
will be confirmed with better accuracy, values of
$(\delta^d_{23})_{AB}$ close to the naive expectations could be even
favoured by experiments, giving an indirect hint of SUSY in $B$
physics.

\section{Analysis}
We now describe in detail the analysis we carried out. Preliminary
results had been already presented at
ICHEP02~\cite{Silvestrini:2002sm}. We aim at determining the allowed
regions in the SUSY parameter space governing $b \to s$ transitions,
studying the correlations among different observables and pointing out
possible signals of SUSY. The constraints on parameter space come
from:
\begin{enumerate}
\item The BR$(B \to X_s \gamma)=(3.29 \pm 0.34)\times
  10^{-4}$ (experimental results as reported in~\cite{Stocchi:2002yi},
  rescaled according to ref.~\cite{Gambino:2001ew}). 
  
\item The CP asymmetry $A_{CP}(B \to X_s \gamma)=-0.02 \pm
  0.04$~\cite{Stocchi:2002yi}.

\item The BR$(B \to X_s \ell^+ \ell^-)=(6.1 \pm 1.4 \pm 1.3)\times
  10^{-6}$~\cite{Stocchi:2002yi}.

\item The lower bound on the $B_s - \bar B_s$ mass difference $\Delta
  M_{B_s} > 14.4$ ps$^{-1}$ \cite{Stocchi:2002yi}.
\end{enumerate}
We have also considered BR's and CP asymmetries for $B \to K \pi$.
For $B \to \phi K_s$, we have studied the BR and the
coefficients $C_{\phi K}$ and $S_{\phi K}$ of cosine and sine terms in
the time-dependent CP asymmetry. Due to theoretical and experimental
uncertainties, these processes are less effective as a constraint on
SUSY, as we discuss in the following.

Concerning $B_s - \bar B_s$ mixing, we closely follow the treatment of
$B_d - \bar B_d$ mixing in ref.~\cite{Becirevic:2001jj}, where all the
relevant formulae for matching conditions~\cite{Gabbiani:1996hi}, NLO
QCD evolution~\cite{Ciuchini,Buras:2001ra} and hadronic matrix
elements~\cite{Becirevic:2001yv} can be found (with the obvious
replacement $d \to s$), as well as a detailed discussion of the
uncertainties of this kind of analysis. We include SM and
gluino-mediated SUSY contributions, which are expected to be dominant
for large mass insertions. However, in particular cases, such as large
$\mu \tan \beta$ scenarios, chargino exchange can also be important.
See ref.~\cite{Gabrielli:2002fr} for a discussion of the impact of
chargino-mediated contributions in $\Delta B=2$ processes.

For $\Delta B=1$ transitions, the matching conditions for four-quark
operators are given in ref.~\cite{match}, and the NLO QCD evolution
can be obtained from the anomalous dimensions given in
ref.~\cite{NLODB1}. The NLO anomalous dimensions for magnetic and
chromomagnetic operators can be found in ref.~\cite{Chetyrkin:1996vx}.
The matching conditions including semileptonic operators are given in
ref.~\cite{Cho:1996we} and the anomalous dimensions were given in
ref.~\cite{Misiak:bc}. NLO QCD matrix elements and matching conditions
to the Standard Model for $B \to X_s \gamma$ can be found in
ref.~\cite{bsgme}.

Concerning $B \to K \pi$ and $B \to \phi K_s$ decays, we adopt BBNS
factorization~\cite{BBNS}, with the caveats on $\Lambda/M_b$
corrections raised in ref.~\cite{Charming}. The importance of
power-suppressed contributions in $B \to K \pi$ decays is widely
recognized. However, it is a matter of debate whether these
corrections come mainly from the matrix elements of penguin operators
or from penguin contractions of current-current operators containing
charm quarks (the charming penguins of ref.~\cite{Ciuchini:1997hb}).
These contributions, having the same quantum numbers, cannot be
disentangled using experimental data. Indeed, present data can be
reproduced using either mechanism. SUSY only affects the coefficients
of penguin operators, leaving current-current operators unmodified.
Therefore, the sensitivity to SUSY effects is expected to be much
lower in the presence of charming penguins. For this reason, in order
to maximize the sensitivity to SUSY contributions, we use the BBNS
treatment of power-suppressed terms in this paper. However, in the
spirit of the criticism of ref.~\cite{Charming}, we let the
annihilation parameter $\rho_A$ of the BBNS model vary between $0$ and
$8$.

Another source of potentially large SUSY effects in $B \to K \pi$ and
$B \to \phi K_s$ decays is the contribution of the chromomagnetic
operator. As pointed out in ref.~\cite{c8g}, this operator can be
substantially enhanced by SUSY without spoiling the experimental
constraints from $B\to X_s\gamma$.  Also here the hadronic
uncertainties can be large. In fact, this term is generated as an
${\cal O}(\alpha_s)$ correction to the leading power amplitude. For
this reason, the one-loop proof of factorization does not apply to
this term. In any case, power-suppressed contributions may be
numerically of the same size.

We have checked explicitly that, given the large hadronic
uncertainties, $B \to K \pi$ modes give no significant constraints on
the $\delta$'s. The time-dependent asymmetry in $B \to \phi K_s$,
instead, is sensitive to the SUSY parameters. This sensitivity is more
pronounced in the case of $LR$ insertions. However, this comes from the
contribution of the chromomagnetic operator, via the matrix element
that, as already mentioned, has large uncertainties.

We performed a MonteCarlo analysis, generating weighted random
configurations of input parameters (see ref.~\cite{Ciuchini:2000de}
for details of this procedure) and computing for each configuration
the processes listed above. We study the clustering induced by the
contraints on various observables and parameters, assuming that each
unconstrained $\delta_{23}^d$ fills uniformly a square $(-1\dots 1$,
$-1\dots 1)$ in the complex plane. The ranges of CKM parameters have
been taken from the UT fit~\cite{ckm2} ($\bar \rho=0.178 \pm
0.046$, $\bar \eta=0.341 \pm 0.028$), and hadronic parameter ranges
are as given in
refs.~\cite{Stocchi:2002yi,Gambino:2001ew,Becirevic:2001yv,Charming}.

Concerning SUSY parameters, we fix $m_{\tilde q}=m_{\tilde g}=350$ GeV
and consider different possibilities for the mass insertions. In
addition to studying single insertions, we also examine the
effects of the left-right symmetric case
$(\delta^d_{23})_{LL}=(\delta^d_{23})_{RR}$ and of
$(\delta^d_{23})_{RR}=(\delta^d_{23})_{LR}$ inspired by large $RR$
mixing at large tan$\beta$~\cite{Harnik:2002vs}. 

The gluino-mediated $b \to s$ transitions in the MSSM had already been
investigated by several
authors~\cite{Bertolini:1987pk}--\cite{Causse:2002mu} before the
announcement of $S_{\phi K}$ negative and have been vigorously
reassessed~\cite{Harnik:2002vs,Hiller:2002ci}--\cite{Khalil:2002fm}
after such results were announced last Summer. In particular in the
works of refs.~\cite{Kane:2002sp,Khalil:2002fm} the correlation
between $B \to \phi K_s$ and $B_s - \bar B_s$ mixing has been
investigated making use of the mass insertion approximation. In our
present work we largely improve at the level of accuracy with the
inclusion of NLO QCD corrections and lattice QCD hadronic matrix
elements and also in the correlation of $b \to s$ related processes
with the selection of the $\Delta B=1$ and $\Delta B=2$ phenomena
outlined above. Indeed, for instance the inclusion of the process
$B \to X_s \ell^+ \ell^-$ leads to further constraints which were not
included in such previous analyses. As for the evaluation of $B \to K
\pi$ and $B \to \phi K_s$, ref.~\cite{Kane:2002sp} adopts the BBNS
factorization, but without discussing the possibly large $\Lambda/M_b$
corrections. This may be the source of some relevant quantitative
difference on the $RR$ contributions to $BR(B \to \phi K_s)$ with
respect to refs.~\cite{Kane:2002sp,Khalil:2002fm}, as we will detail
in next Section. As for the analysis of ref.~\cite{Harnik:2002vs},
this is performed in the mass eigenstate basis taking a specific down
squark mass matrix (as suggested in SUSY GUT's where the large
neutrino mixing is linked to a large $\tilde b_R \to \tilde s_R$
mixing). Comparing the results of our cases of $RR$ dominance and
$RR=RL$ dominance with their results we
find some discrepancy in particular in the case of large
$(\delta_{23}^d)_{RR}$ (see below). Once again a potential source of
discrepancy in constraining the $\delta^d_{23}$'s from $A_{CP}(B\to
\phi K_s)$ is represented by the delicate evaluation of the matrix
elements of the chromo-dipole operators.

\begin{figure}
  \begin{center}
    \begin{tabular}{c c}
      \includegraphics[width=0.48\textwidth]{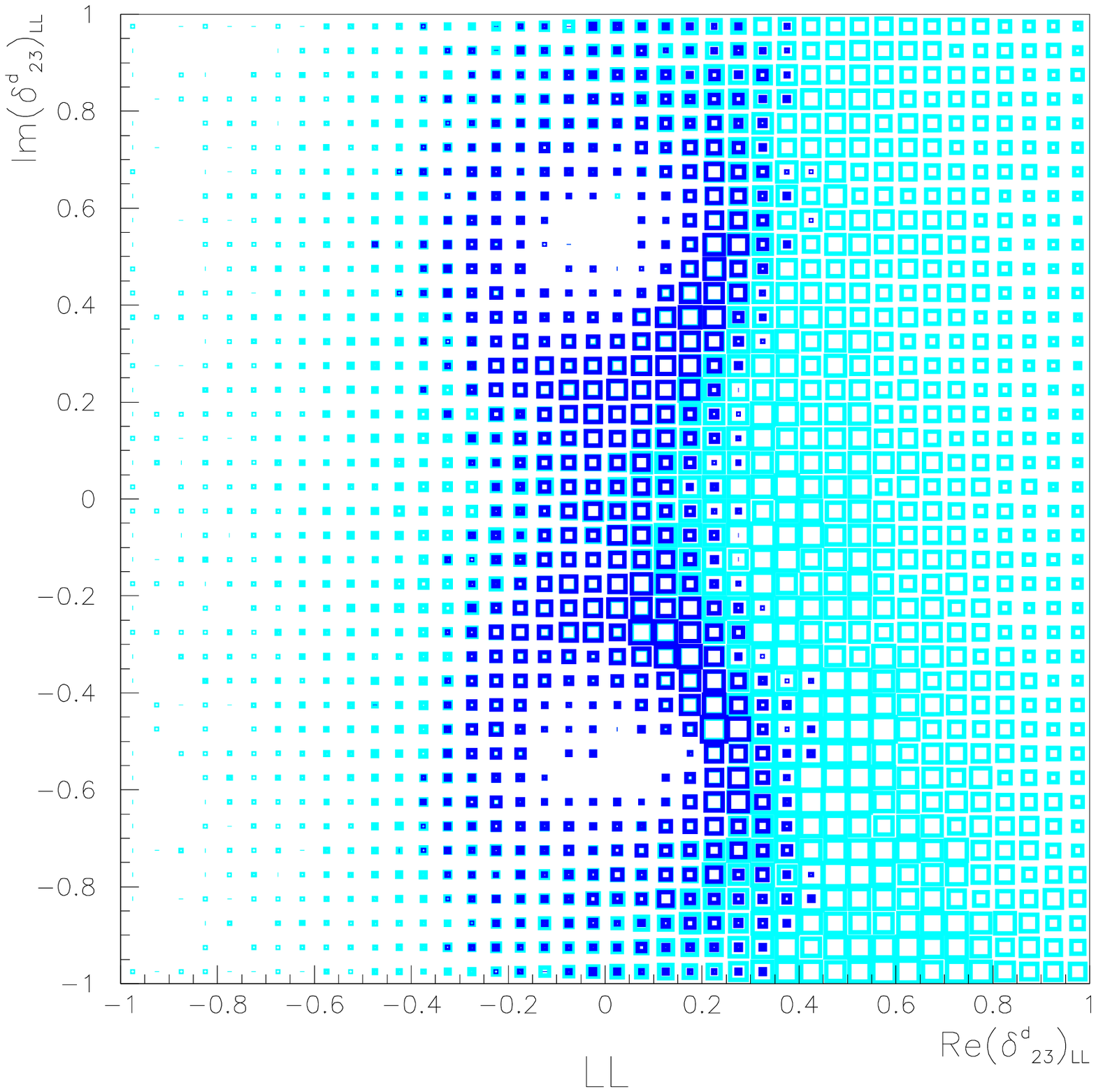} &
      \includegraphics[width=0.48\textwidth]{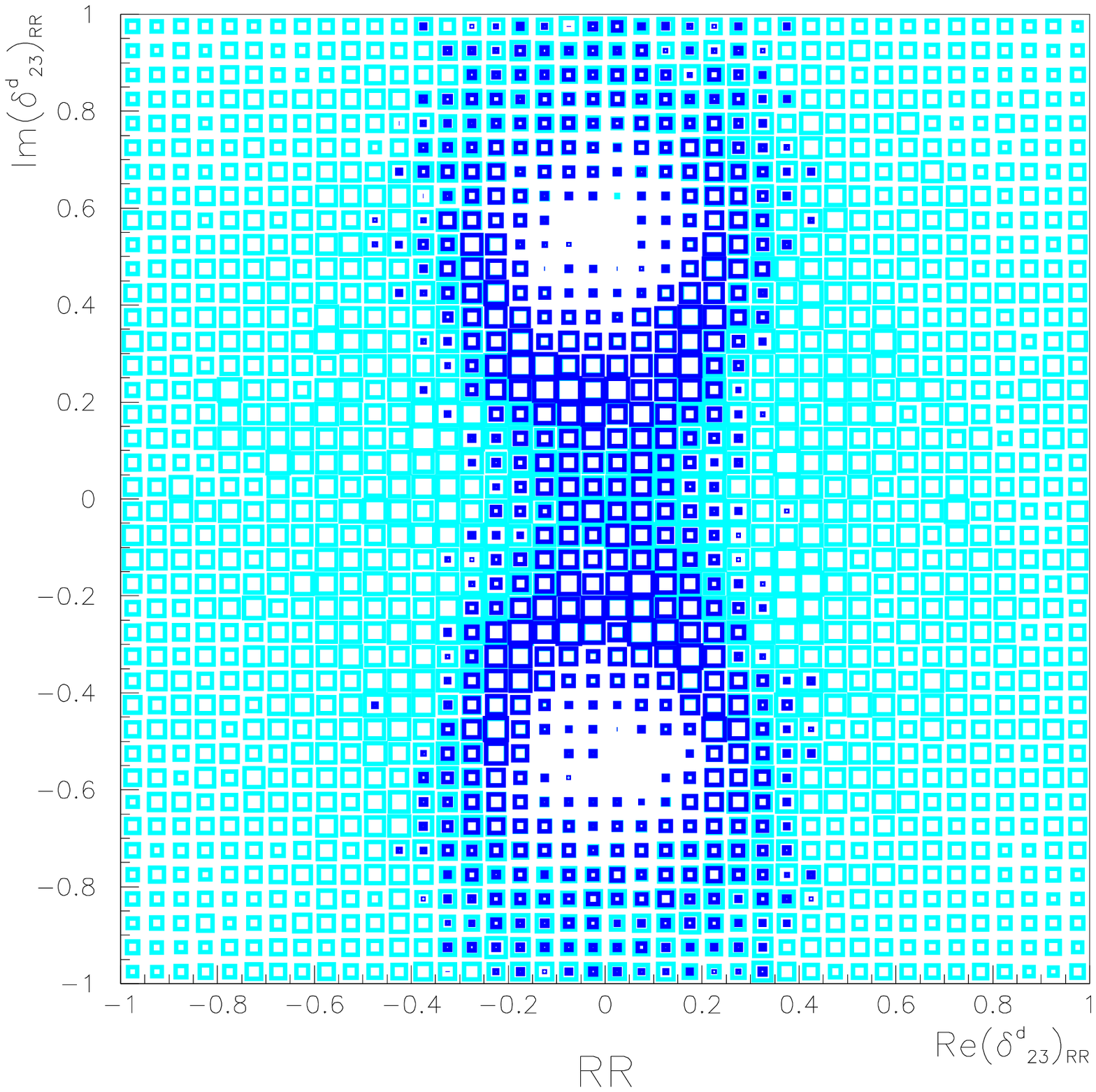} \\
      \includegraphics[width=0.48\textwidth]{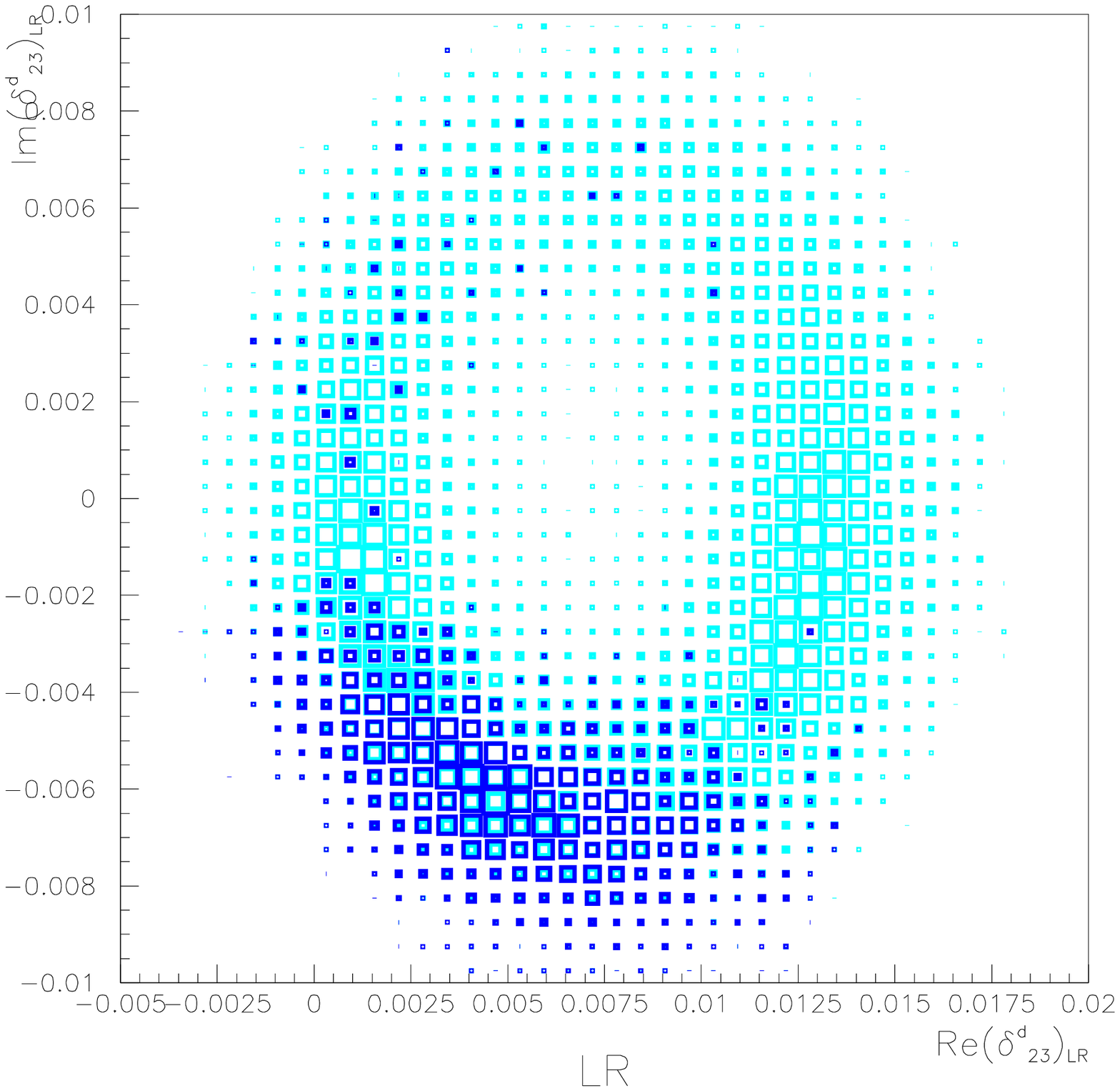} &
      \includegraphics[width=0.48\textwidth]{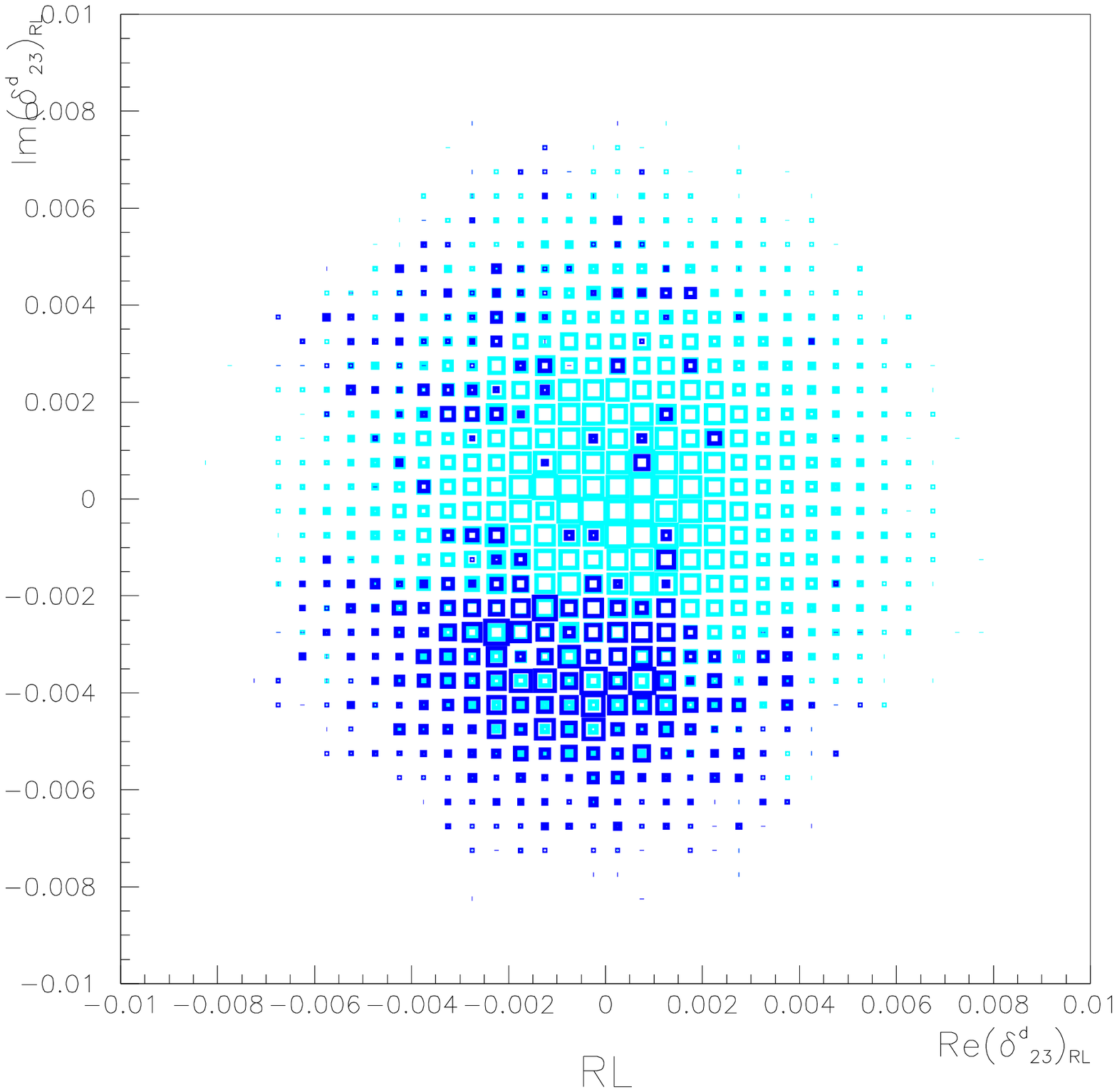} \\ 
    \end{tabular}
  \end{center}
  \caption{Allowed regions in the  
    Re$(\delta^d_{23})_{AB}$--Im$(\delta^d_{23})_{AB}$ space for
    $m_{\tilde q}=m_{\tilde g}=350$ GeV and $AB=(LL,RR,LR,RL)$. 
    Constraints from $BR(B\to
    X_s\gamma)$, $A_{CP}(B\to
    X_s\gamma)$, $BR(B\to
    X_sl^+l^-)$ and the lower bound on $\Delta M_s$ have been used.
    The darker
    regions are selected imposing the further constraint $\Delta
    m_s<20$ ps$^{-1}$ for
    $LL$ and $RR$ insertions and $S_{\phi K}<0$ for
    $LR$ and $RL$ insertions.}
  \label{fig:ranges1}
\end{figure}

\begin{figure}[t]
  \begin{center}
    \begin{tabular}{c c}
      \includegraphics[width=0.48\textwidth]{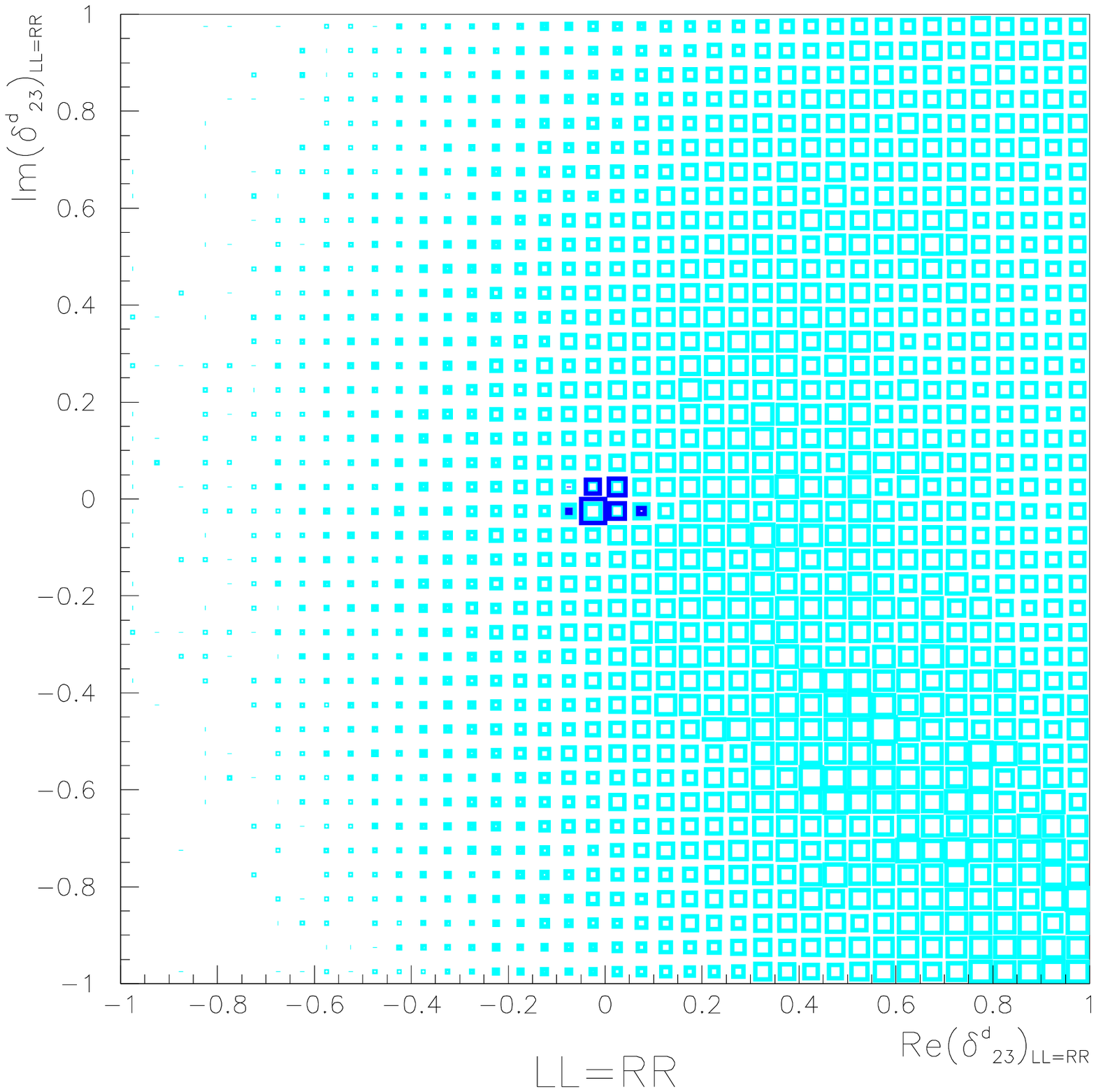} &
      \includegraphics[width=0.48\textwidth]{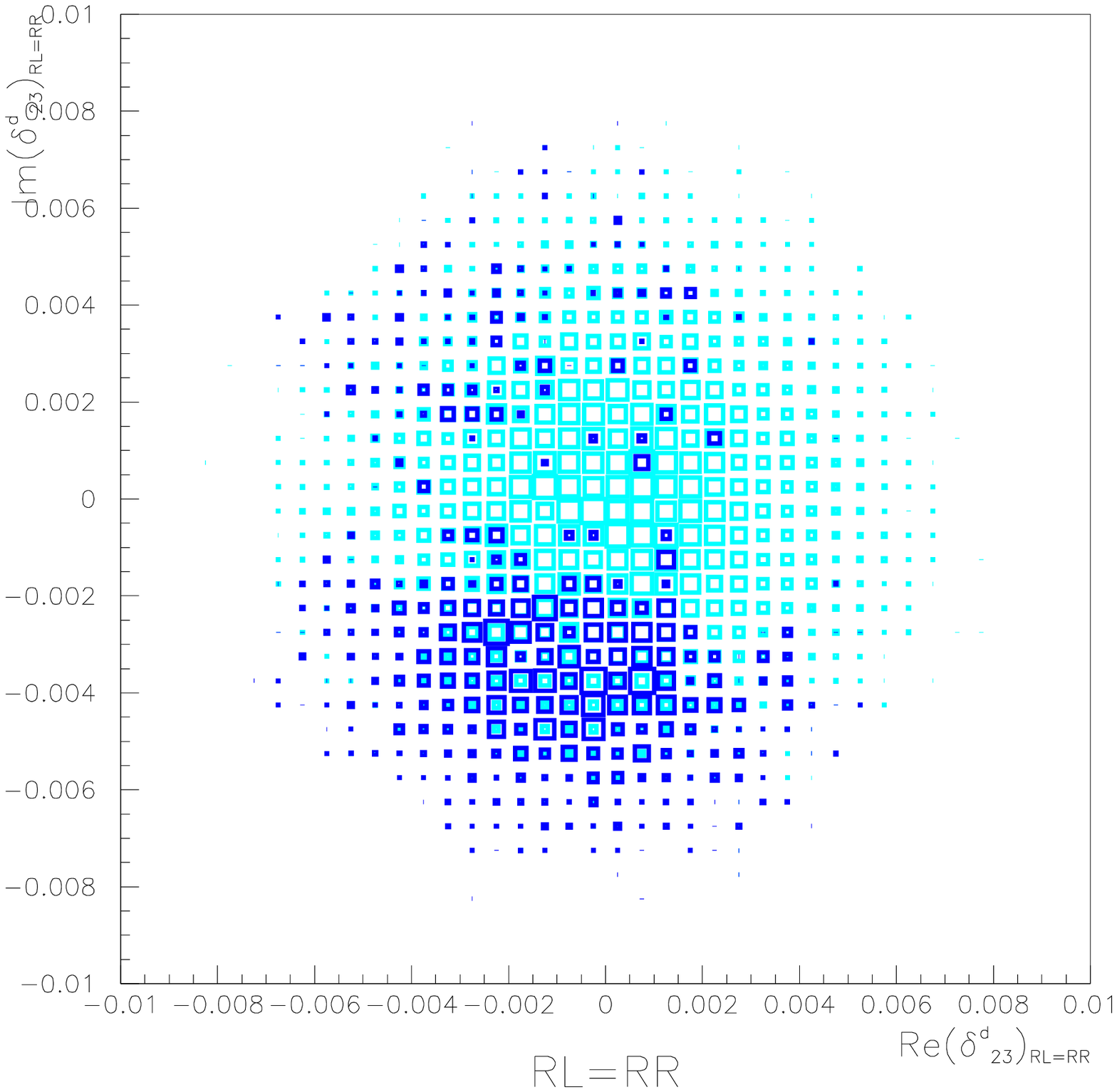} \\
      \\
    \end{tabular}
  \end{center}
  \caption{Allowed regions in the  
    Re$(\delta^d_{23})_{AB}$--Im$(\delta^d_{23})_{AB}$ space for
    $m_{\tilde q}=m_{\tilde g}=350$ GeV for double mass insertions
    $LL=RR$ (left) and $RL=RR$ (right). Constraints from $BR(B\to
    X_s\gamma)$, $A_{CP}(B\to X_s\gamma)$, $BR(B\to X_sl^+l^-)$ and
    the lower bound on $\Delta M_s$ have been used. The darker
    regions are selected imposing the further constraint $\Delta
    m_s<20$ ps$^{-1}$ for $LL=RR$ insertions and $S_{\phi K}<0$
    for $RL=RR$ insertions.}
  \label{fig:ranges2}
\end{figure}

In fig.~\ref{fig:ranges1} we display the clustering of events in the
Re$(\delta^d_{23})_{AB}$--Im$(\delta^d_{23})_{AB}$ plane in the single
insertion case. Here and in the following plots, larger boxes
correspond to larger numbers of weighted events. Constraints from
$BR(B\to X_s\gamma)$, $A_{CP}(B\to X_s\gamma)$, $BR(B\to X_s \ell^+
\ell^-)$ and the lower bound on $\Delta M_s$ have been used, as
discussed above. The darker regions are selected imposing the further
constraint $\Delta M_s<20$~ps$^{-1}$ for $LL$ and $RR$ insertions and
$S_{\phi K}<0$ for $LR$ and $RL$ insertions.  For helicity conserving
insertions, the constraints are of order $1$. A significant reduction
of the allowed region appears if the cut on $\Delta M_s$ is imposed.
The asymmetry of the $LL$ plot is due to the interference with the SM
contribution. In the helicity flipping cases, constraints are of order
$ 1 \times 10^{-2}$. For this values of the parameters,
$\Delta M_s$ is unaffected. We show the effect of requiring $S_{\phi
  K}<0$: it is apparent that a nonvanishing Im$\,\delta_{23}^d$ is
needed to meet this condition.

Fig.~\ref{fig:ranges2} contains the same plots as
fig.~\ref{fig:ranges1} 
in the two cases of double mass insertion that we consider in this
analysis, namely $(\delta^d_{23})_{LL}=(\delta^d_{23})_{RR}$ and
$(\delta^d_{23})_{RL}=(\delta^d_{23})_{RR}$ (see next Section for a
justification of this choice of the double mass insertion cases).
For $(\delta^d_{23})_{LL}=(\delta^d_{23})_{RR}$, the constraints are
still of order $1$, but the contribution to $\Delta M_s$ is huge, due
to the presence of operators with mixed chiralities. This can be seen
from the smallness of the dark region selected by imposing $\Delta
M_s<20$ ps$^{-1}$. As for the double mass insertion with
$(\delta^d_{23})_{RL}=(\delta^d_{23})_{RR}$, by comparison of RHS of
fig.~\ref{fig:ranges2} with the $RL$ case in the lower right side of
fig.~\ref{fig:ranges1}, we argue that both the allowed regions and the
portions of them for which $S_{\phi K}<0$ are very similar in these
two cases. In other words, a double mass insertion with
$(\delta^d_{23})_{RL}=(\delta^d_{23})_{RR}$ does not exhibit
remarkably different features from the case of a single mass insertion
$(\delta^d_{23})_{RL}$, at least at this value of squark and gluino masses. 

\begin{figure}
  \begin{center}
    \begin{tabular}{c c}
      \includegraphics[width=0.48\textwidth]{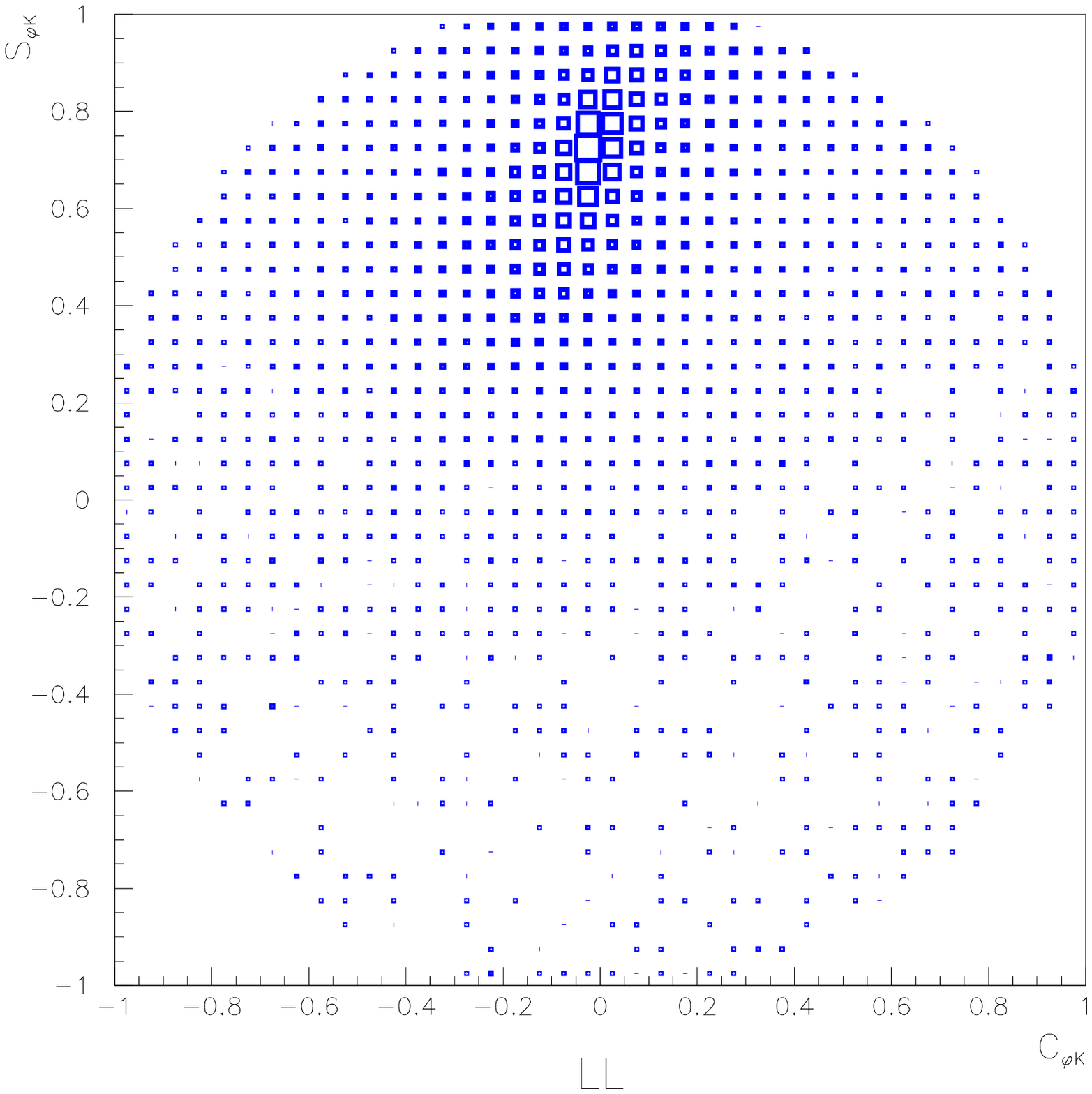} &
      \includegraphics[width=0.48\textwidth]{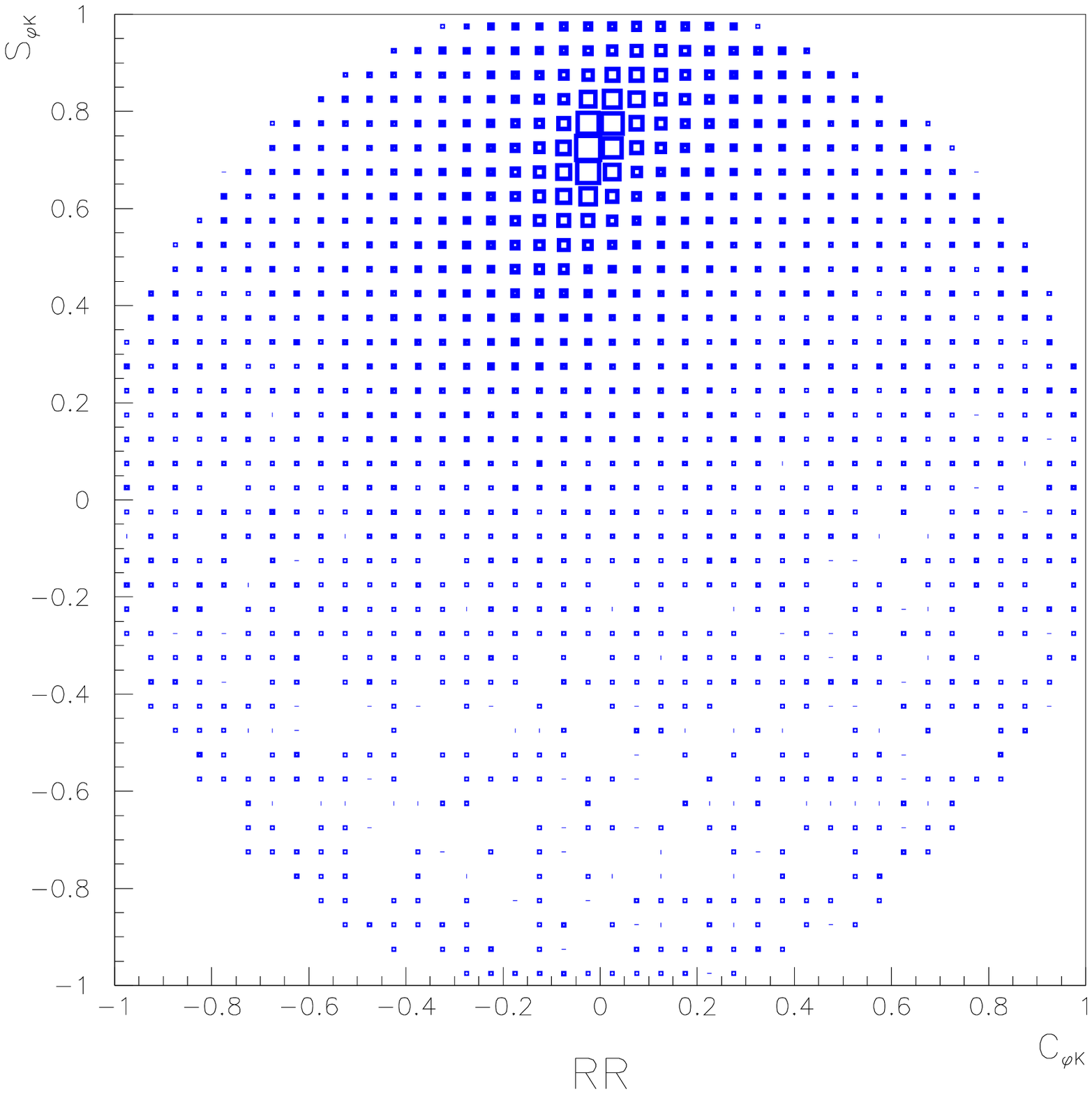} \\
      \includegraphics[width=0.48\textwidth]{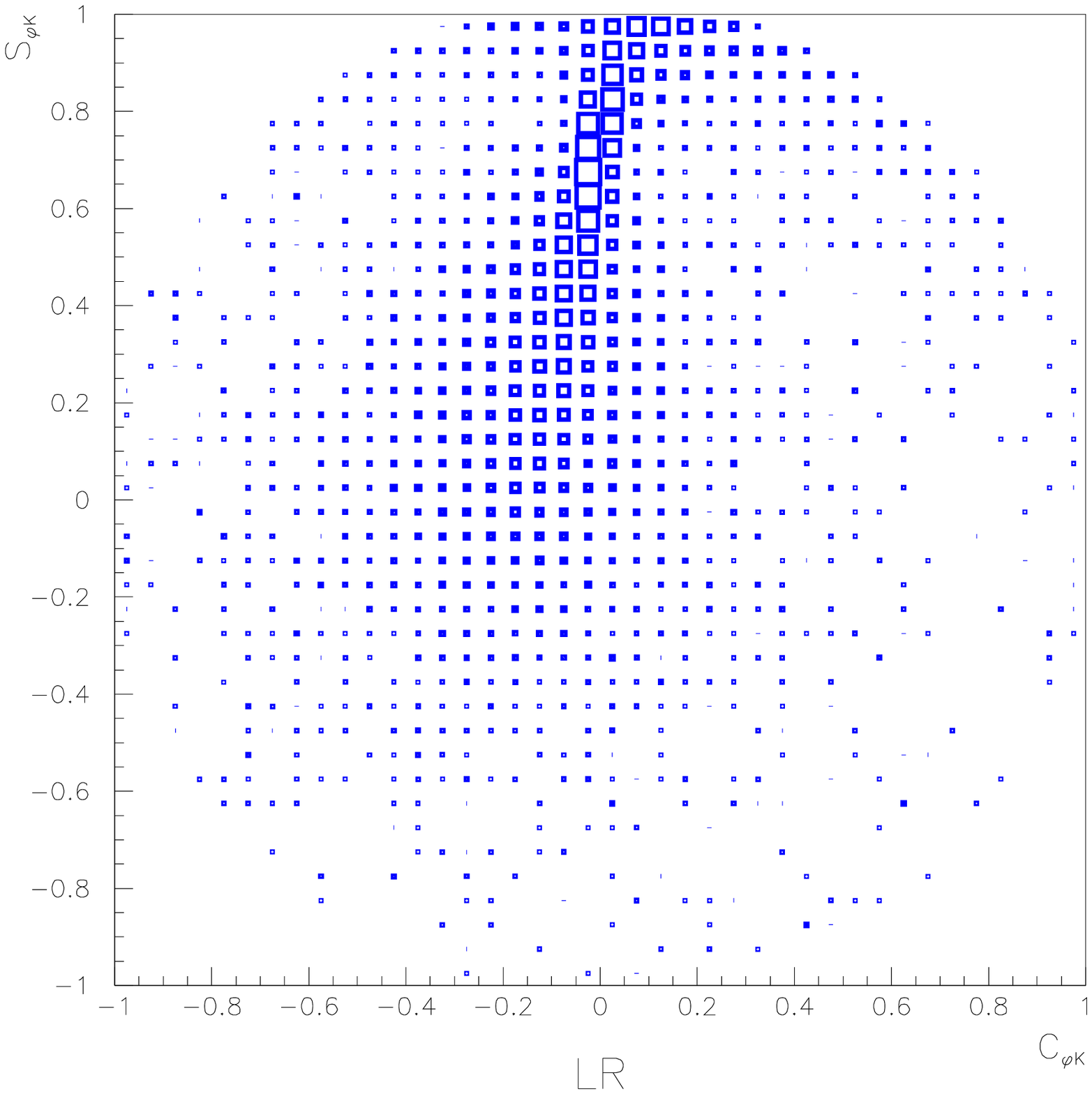} &
      \includegraphics[width=0.48\textwidth]{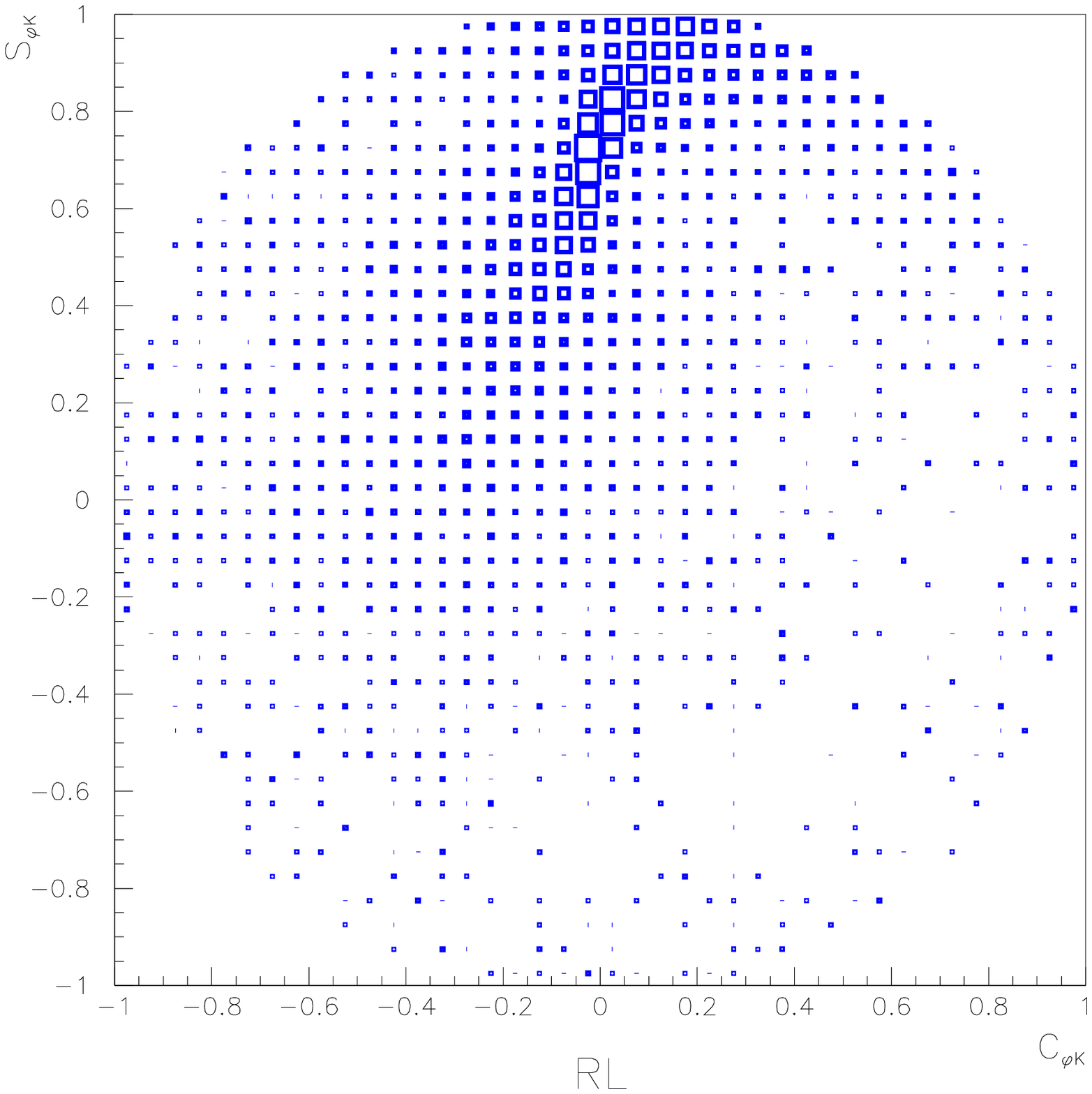} \\ 
    \end{tabular}
  \end{center}
  \caption{Correlations between the sine $(S_{\phi K})$ and cosine
    $(C_{\phi K})$ coefficients of the time-dependent CP asymmetry
    of $B \to \phi K_s$ for
    $m_{\tilde q}=m_{\tilde g}=350$ GeV and various SUSY mass insertions
     $(\delta^d_{23})_{AB}$ with $AB=(LL,RR,LR,RL)$.        
     Constraints from $BR(B\to
    X_s\gamma)$, $A_{CP}(B\to
    X_s\gamma)$, $BR(B\to
    X_sl^+l^-)$ and the lower bound on $\Delta M_s$ have been used.
    }
  \label{fig:sincos1}
\end{figure}
\begin{figure}[t]
  \begin{center}
    \begin{tabular}{c c}
      \includegraphics[width=0.48\textwidth]{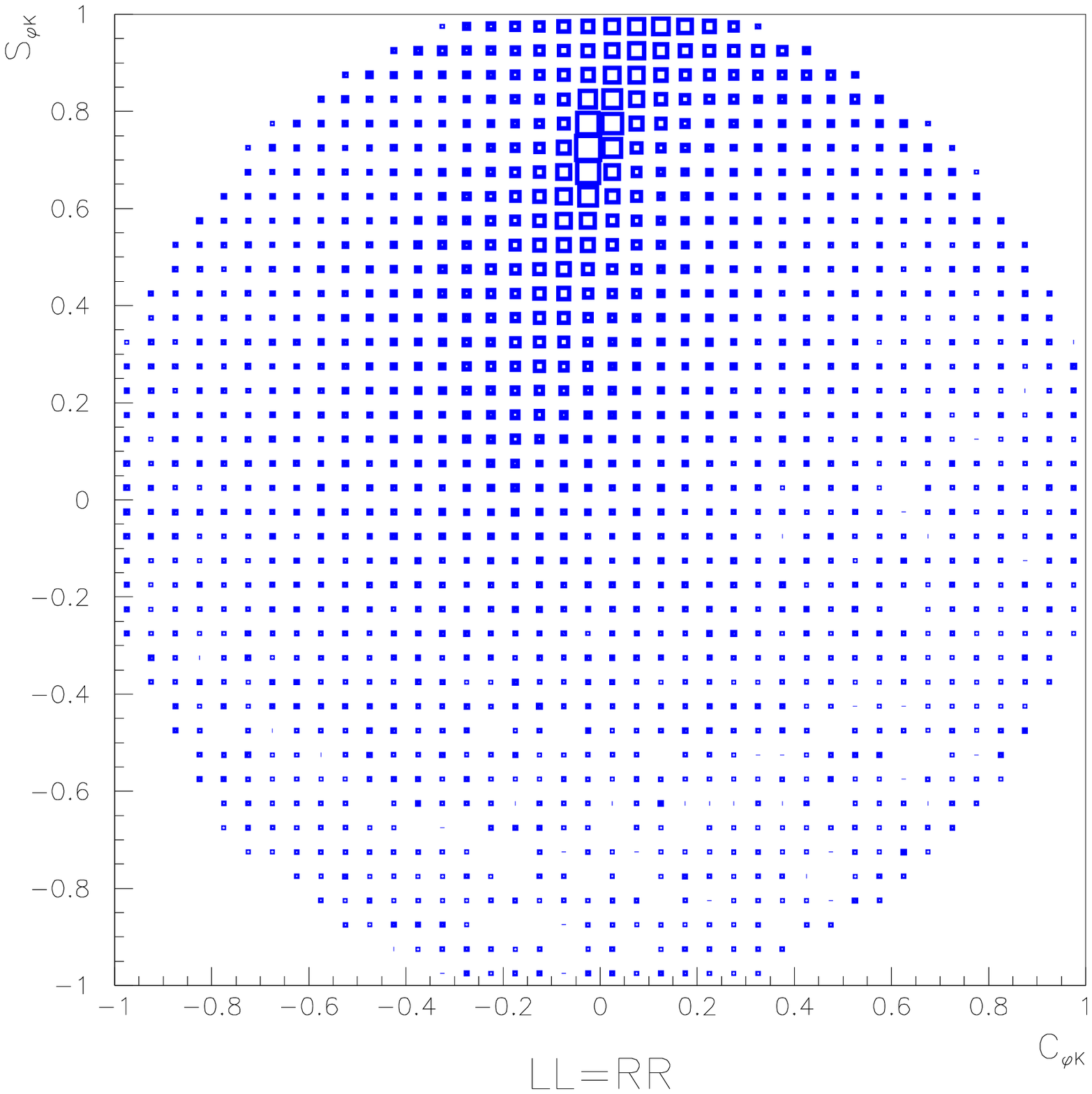} &
      \includegraphics[width=0.48\textwidth]{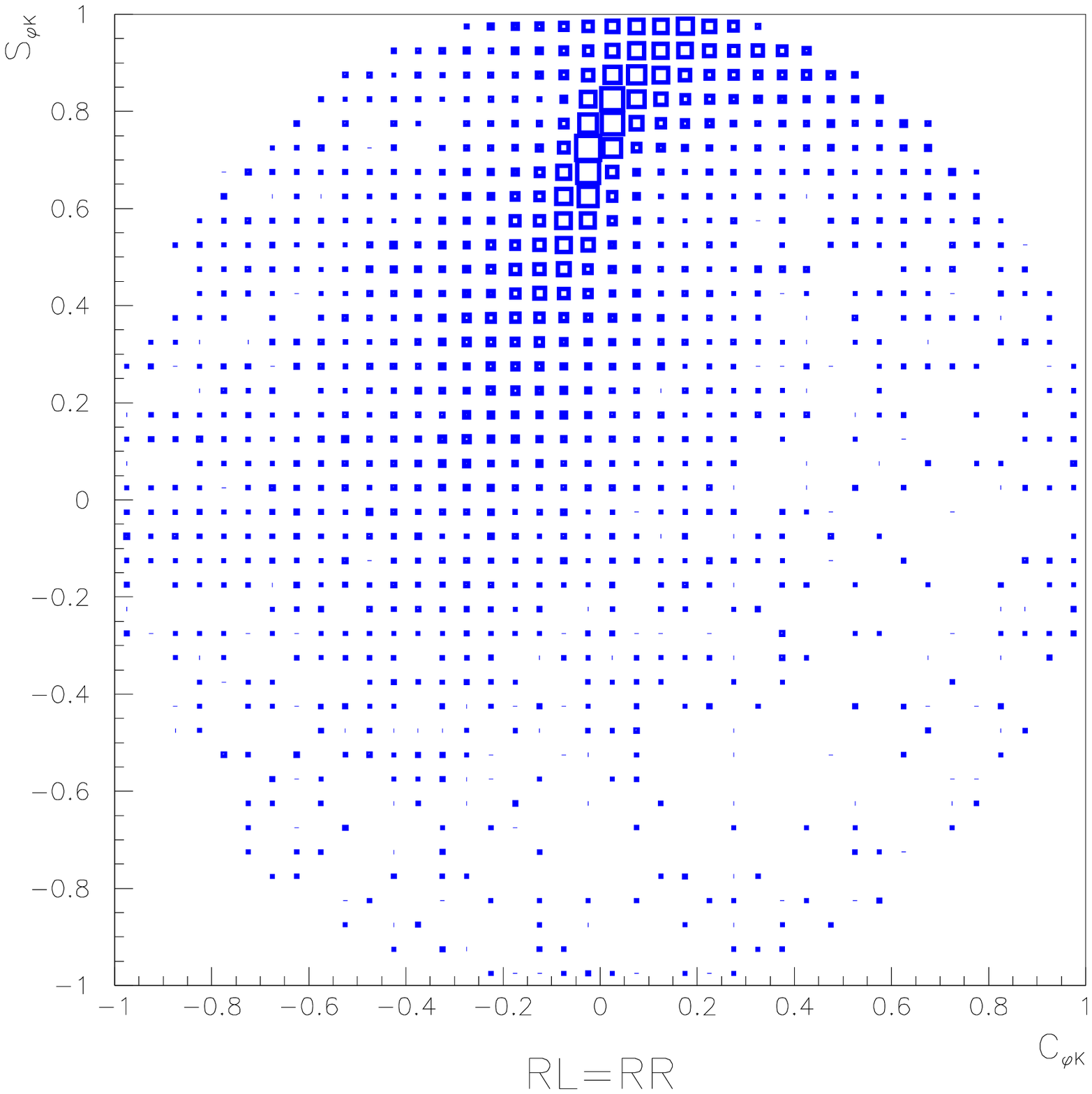} \\
      \\
    \end{tabular}
  \end{center}
  \caption{Same as in fig.~\protect\ref{fig:sincos1} for double 
insertions $LL=RR$ and $RL=RR$.}
  \label{fig:sincos2}
\end{figure}

\begin{figure}
  \begin{center}
    \begin{tabular}{c c}
      \includegraphics[width=0.48\textwidth]{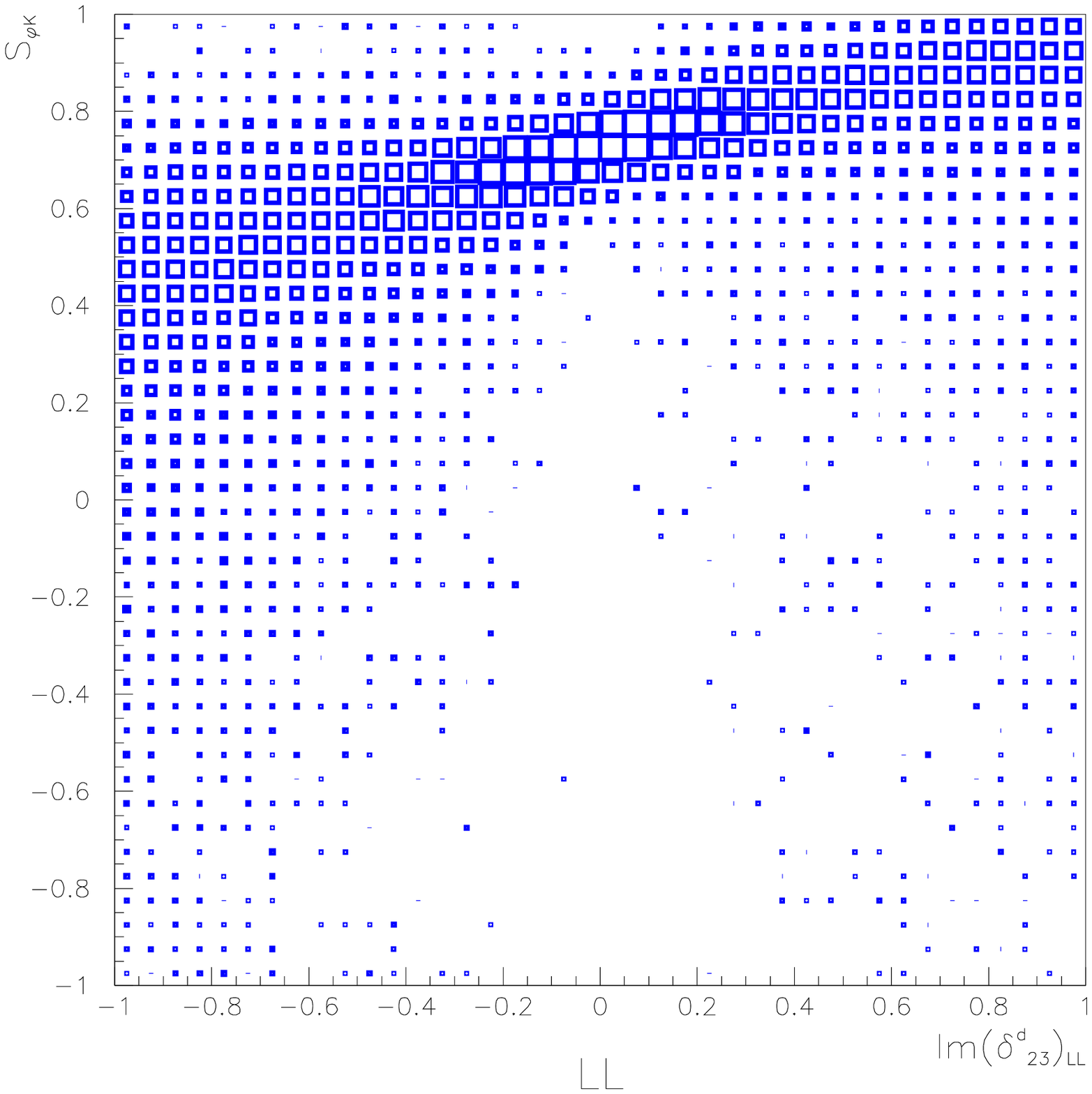} &
      \includegraphics[width=0.48\textwidth]{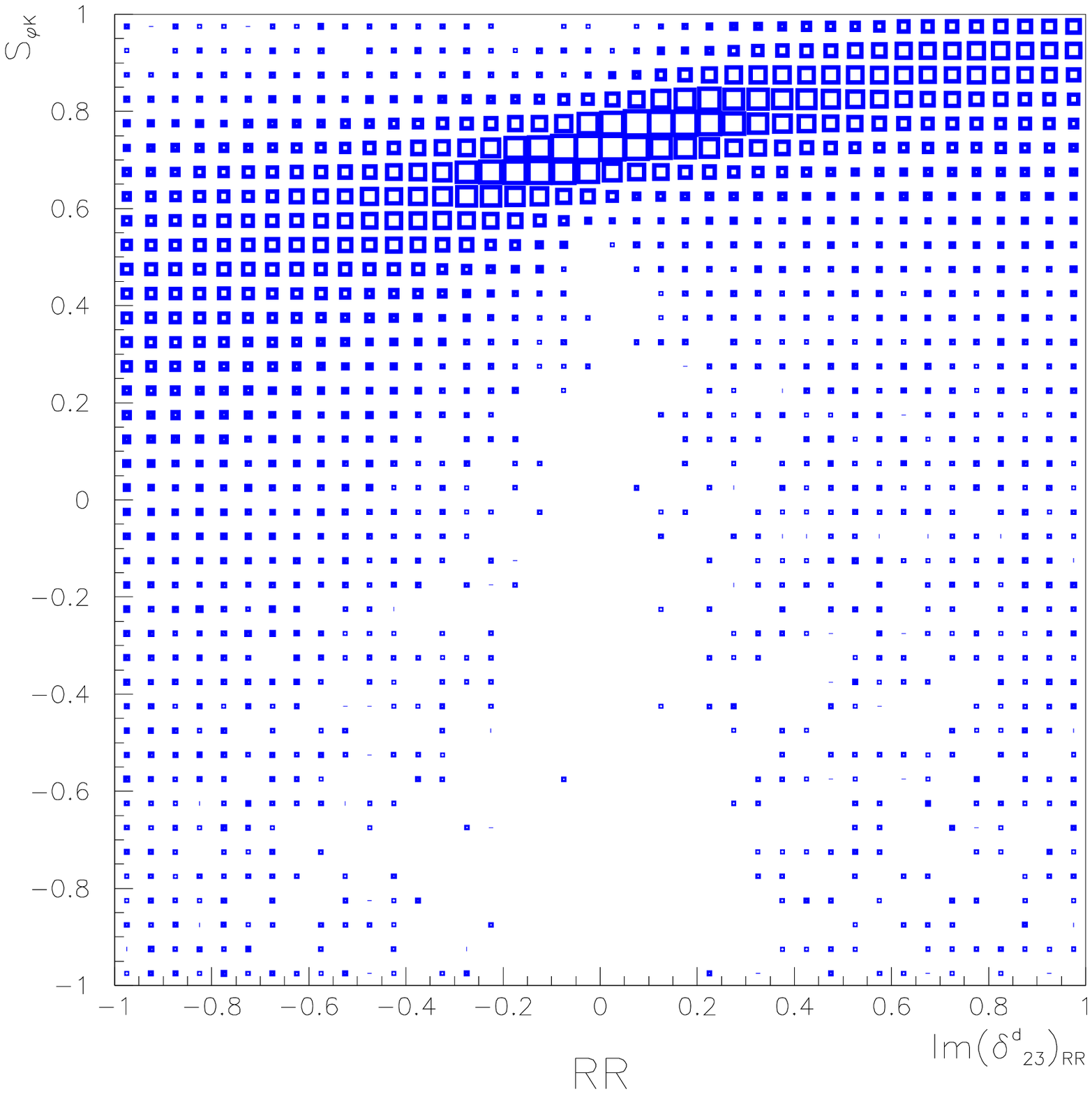} \\
      \includegraphics[width=0.48\textwidth]{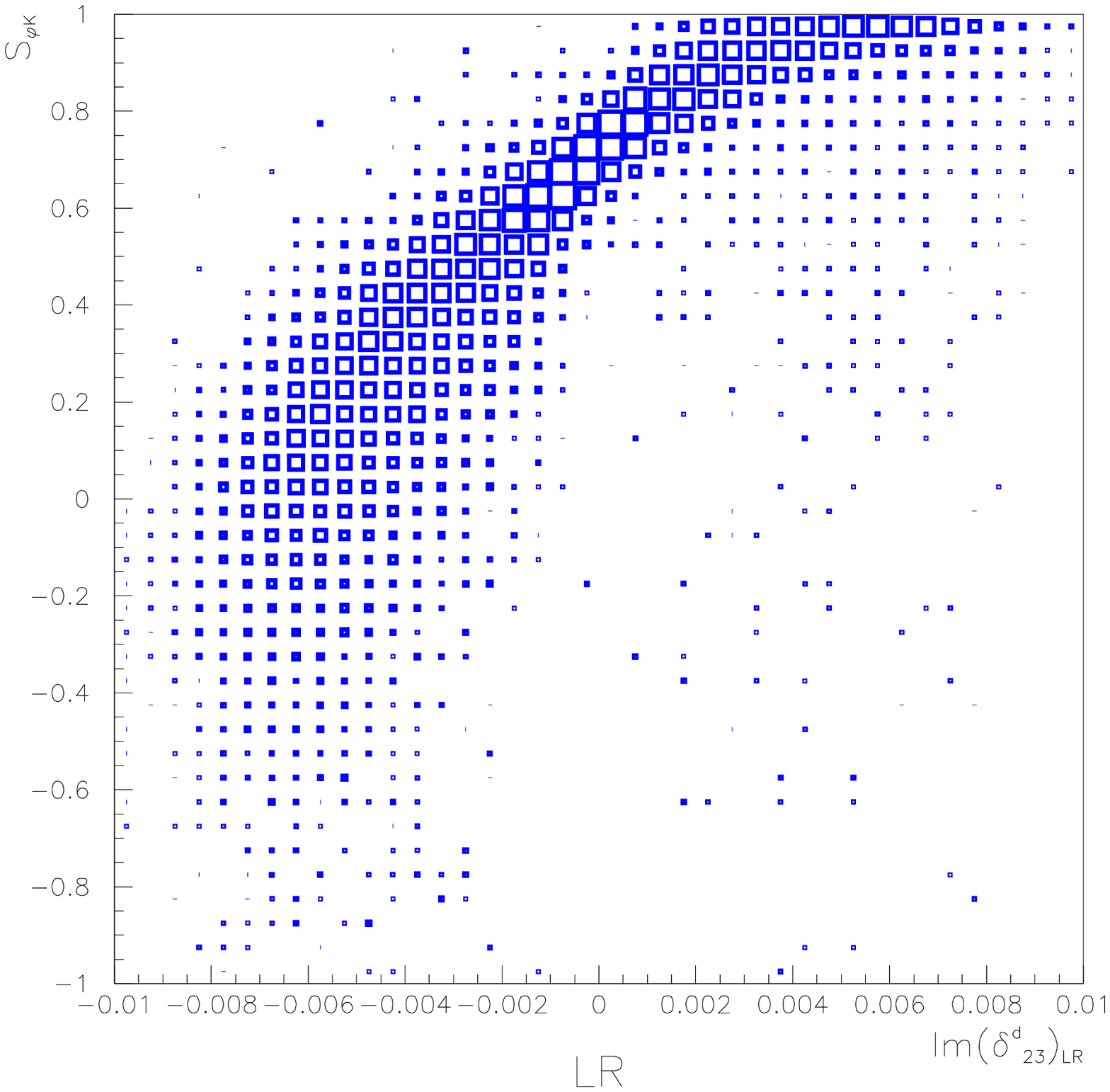} &
      \includegraphics[width=0.48\textwidth]{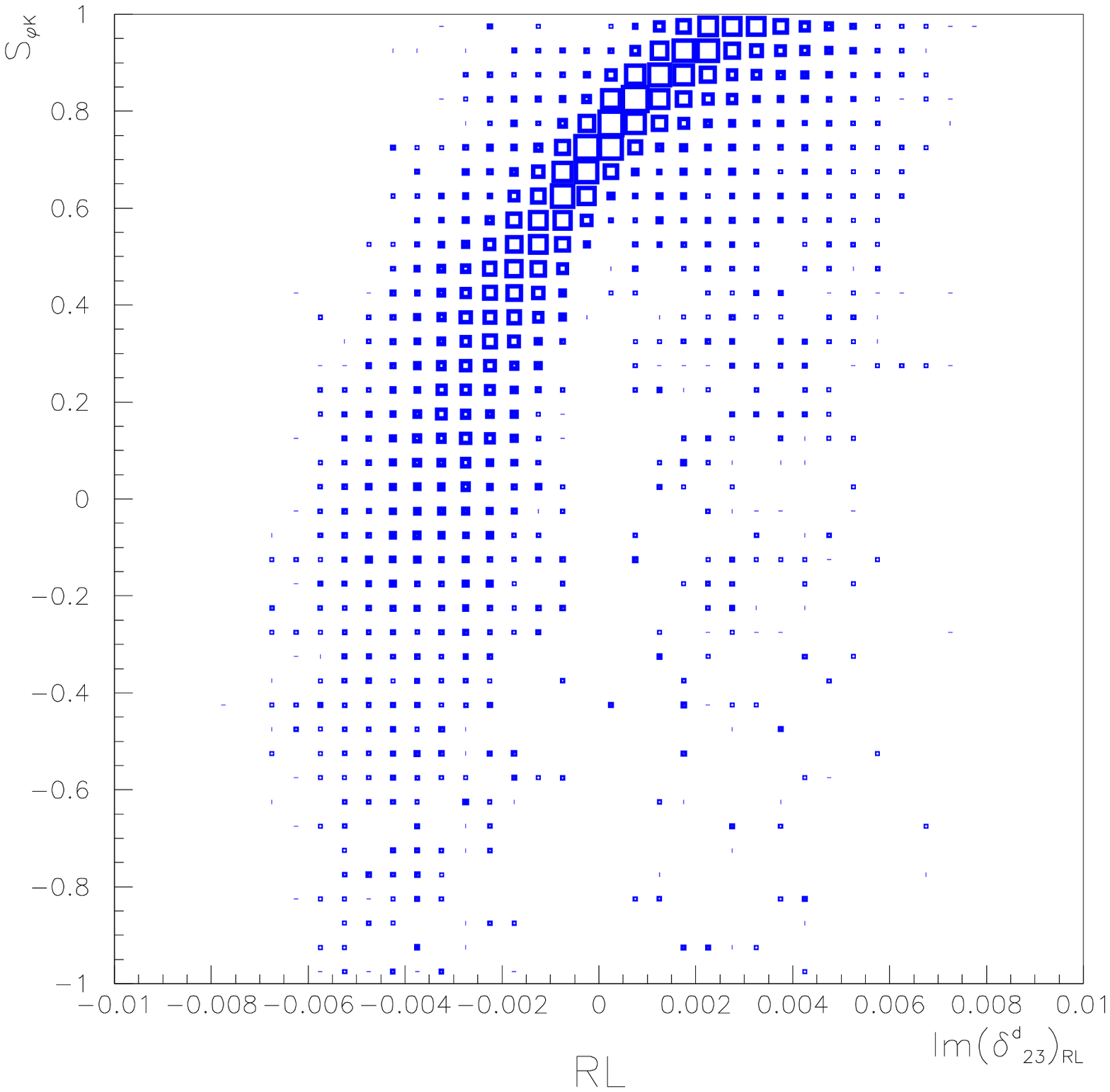} \\ 
    \end{tabular}
  \end{center}
  \caption{Correlations between $S_{\phi K}$ and Im$(\delta^d_{23})_{AB}$ 
    for $m_{\tilde q}=m_{\tilde g}=350$ GeV and   
    $AB=(LL,RR,LR,RL)$.        
    Constraints from $BR(B\to
    X_s\gamma)$, $A_{CP}(B\to
    X_s\gamma)$, $BR(B\to
    X_sl^+l^-)$ and the lower bound on $\Delta M_s$ have been used. 
    }
  \label{fig:sinim1}
\end{figure}

\begin{figure}[t]
  \begin{center}
    \begin{tabular}{c c}
      \includegraphics[width=0.48\textwidth]{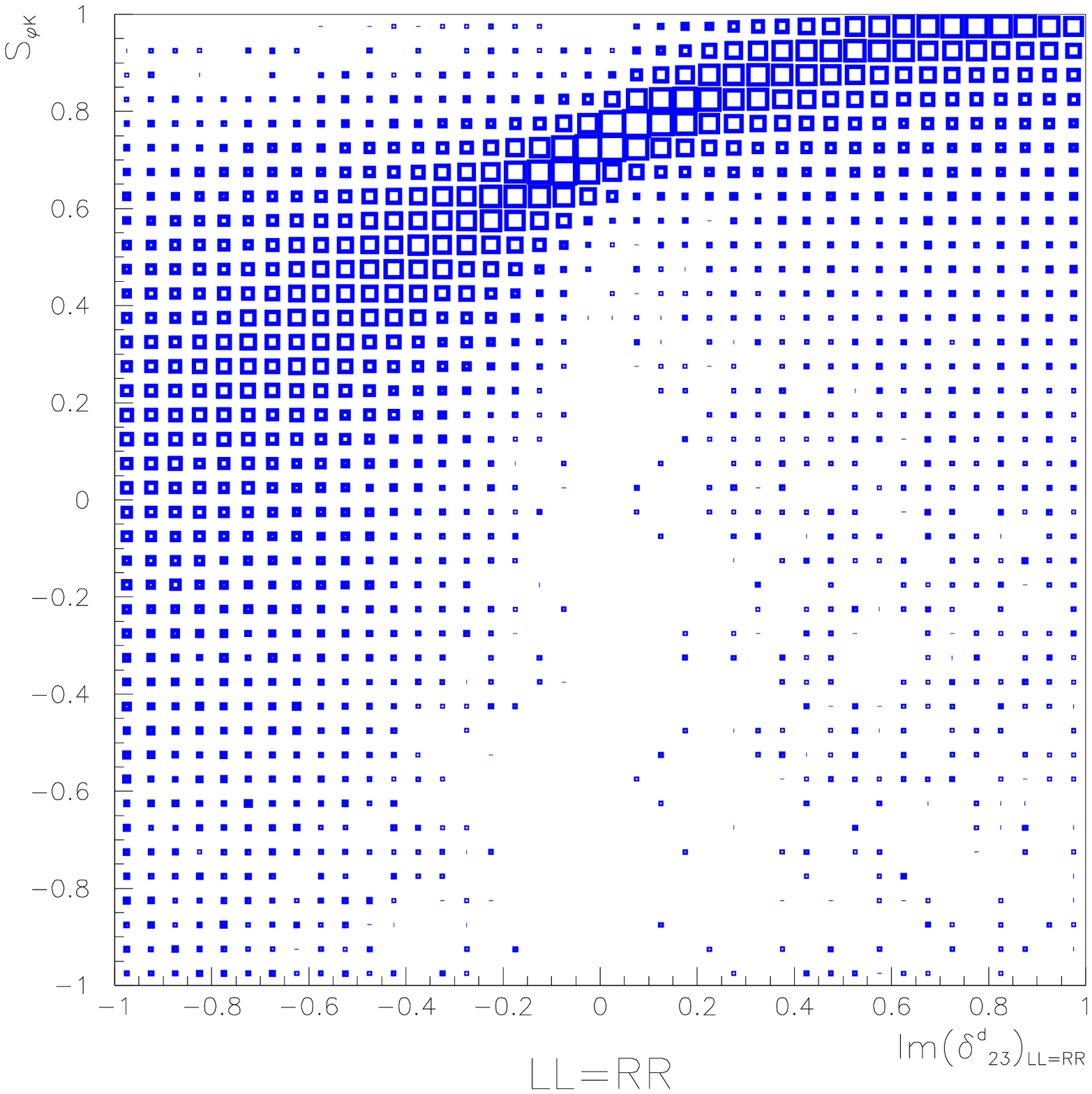} &
      \includegraphics[width=0.48\textwidth]{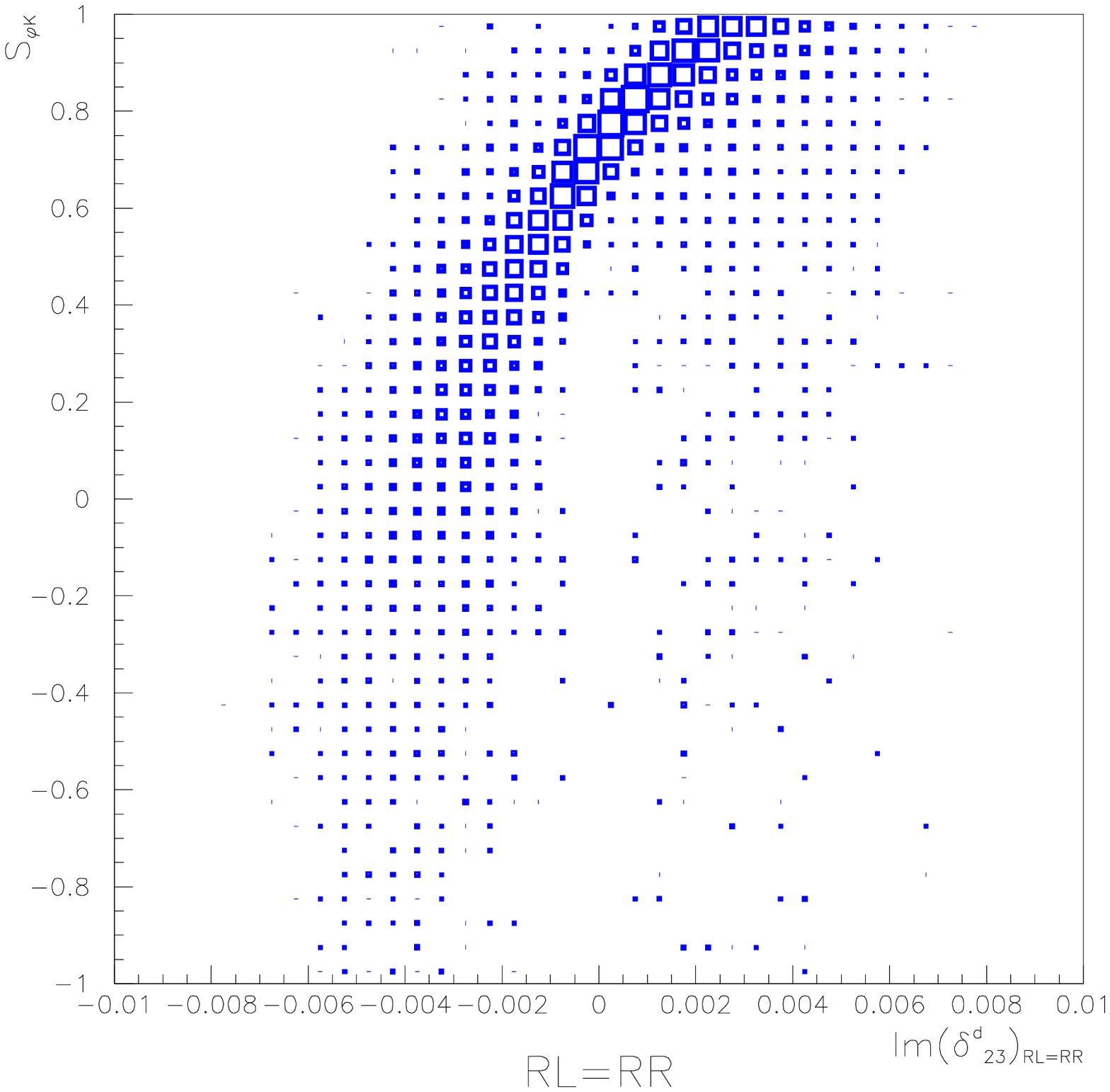} \\
      \\
    \end{tabular}
  \end{center}
  \caption{Same as in fig.~\protect\ref{fig:sinim1} for double 
insertions $LL=RR$ and $RL=RR$.}
  \label{fig:sinim2}
\end{figure}

\begin{figure}
  \begin{center}
    \begin{tabular}{c c}
      \includegraphics[width=0.48\textwidth]{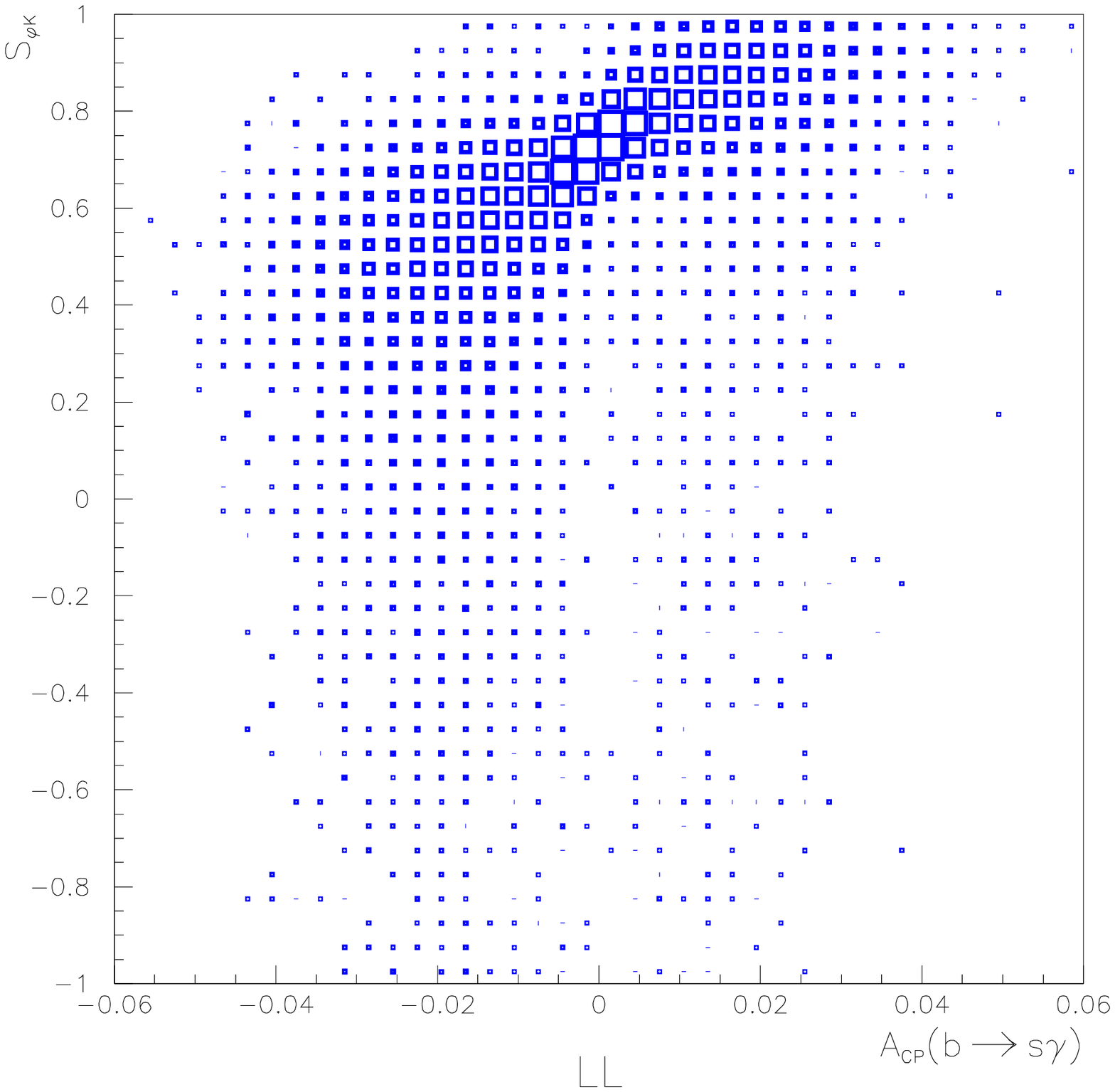} &
      \includegraphics[width=0.48\textwidth]{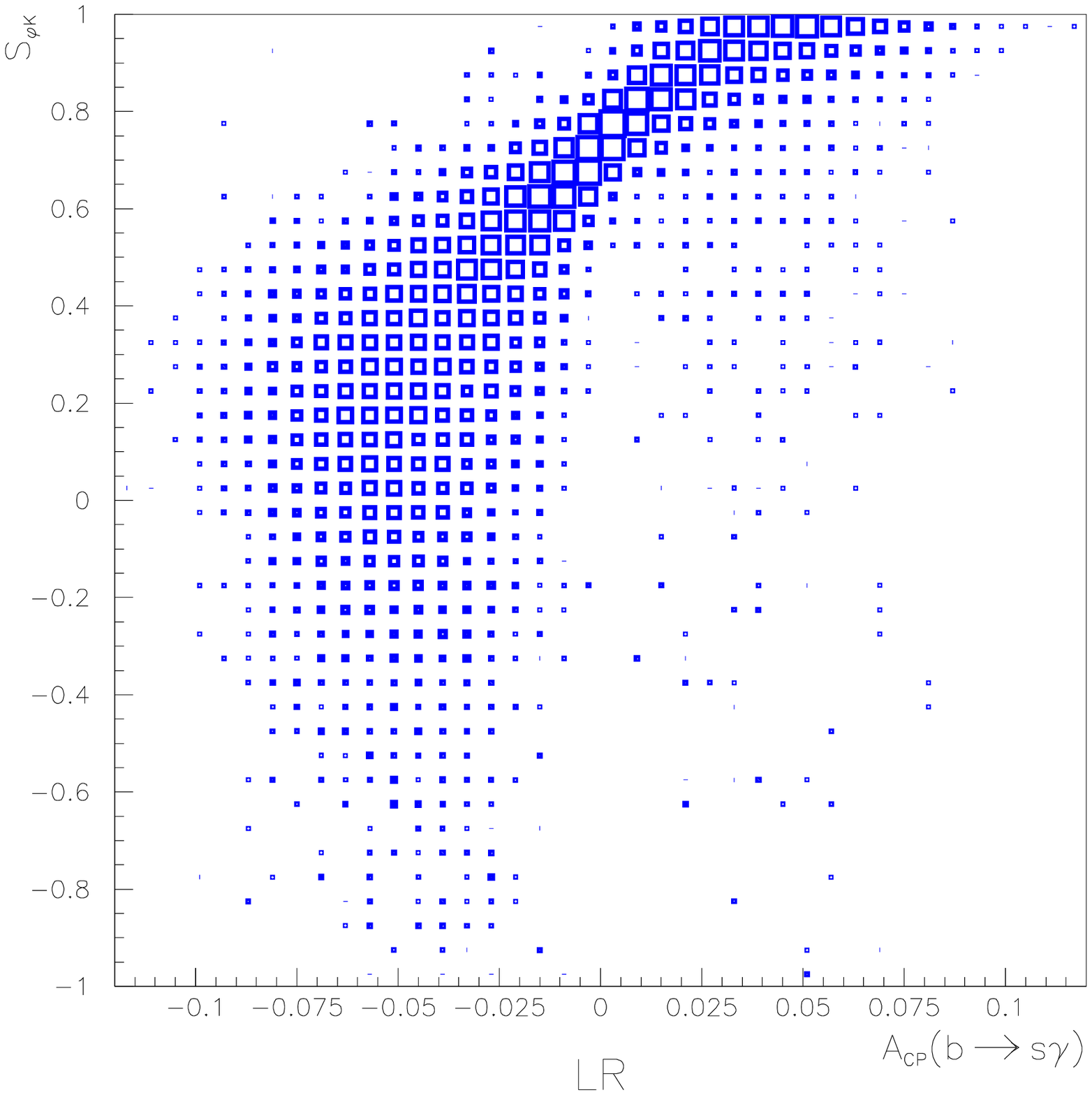} \\
      \includegraphics[width=0.48\textwidth]{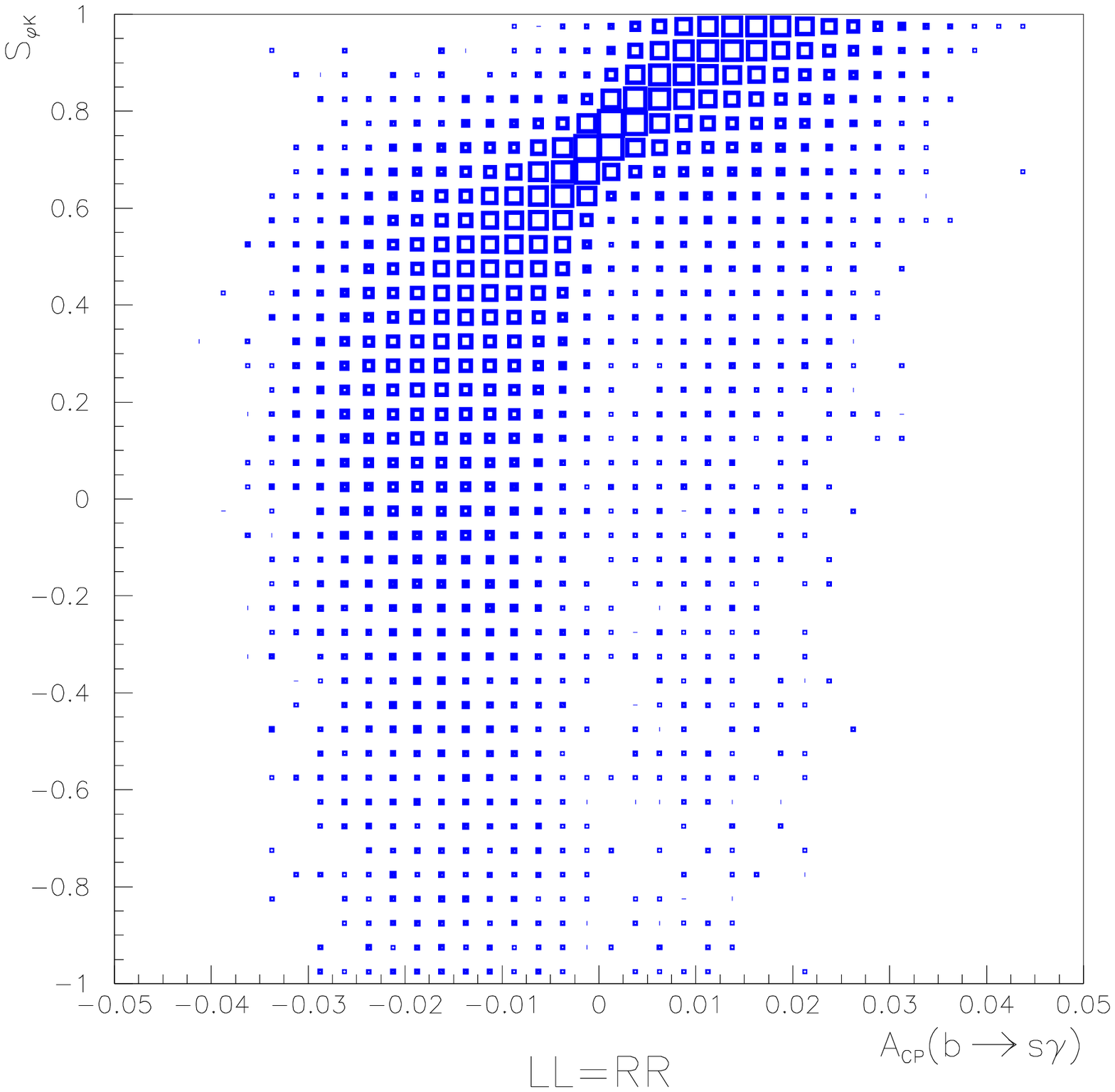} &
      \\
    \end{tabular}
  \end{center}
  \caption{Correlation between $S_{\phi K}$ and $A_{CP}(b\to s\gamma)$
    for various SUSY mass insertions $(\delta^d_{23})_{AB}$ with
    $AB=(LL,LR,LLRR)$. Constraints from $BR(B\to X_s\gamma)$,
    $A_{CP}(B\to X_s\gamma)$, $BR(B\to X_sl^+l^-)$ and the lower bound
    on $\Delta M_s$ have been used.}
  \label{fig:sinabsg}
\end{figure}

In figs.~\ref{fig:sincos1}--\ref{fig:sinabsg}, we study the
correlations between $S_{\phi K}$ and $C_{\phi K}$,
Im$(\delta^d_{23})_{AB}$ and $A_{CP}(B\to X_s\gamma)$ for the various
SUSY insertions considered in the present analysis. In view of the
discussion in the previous Section on the hadronic matrix elements for
$B\to \phi K_s$, the reader should keep in mind that, in all the
results reported in figs.~\ref{fig:sincos1}--\ref{fig:sinabsg}, the
hadronic uncertainties affecting the estimate of $S_{\phi K}$ are not
completely under control. Low values of $S_{\phi K}$ can be more
easily obtained with helicity flipping insertions. A deviation from
the SM value for $S_{\phi K}$ requires a nonvanishing value of
Im$\,(\delta^d_{23})_{AB}$ (see figs.~\ref{fig:sinim1} and
\ref{fig:sinim2}), generating, for those channels in which the SUSY
amplitude can interfere with the SM one, a $A_{CP}(B\to X_s\gamma)$ at
the level of a few percents in the LL and LL=RR cases, and up to the
experimental upper bound in the LR case (see fig.~\ref{fig:sinabsg}).

\section{$b \to s$ transitions: where to look for SUSY?}

We now wish to address the crucial question which naturally arises
after accomplishing the analysis of the constraints on the
$\delta_{23}$ quantities of the previous section: what are the more
promising processes to reveal some indirect signal of low energy SUSY,
among the FCNC ones involving $b \to s$ transitions? As it stands,
this question cannot have a clear-cut answer: indeed, low-energy SUSY
is an ill-defined notion. Rather, sticking as we do here to the MSSM,
one can speak of different MSSM realizations characterized by
different sets of soft breaking terms. For the purpose of the present
discussion, the best way to classify such different ``classes of
MSSM'' is according to the role played by the different
$\delta_{23}^d$'s according to their ``helicities'' $LL$, $RR$, etc.
First we will focus on the case where only one $\delta_{23}^d$
dominates (single mass insertion), while in the final part we will
contemplate the possible coexistence of two sizeable $\delta_{23}^d$'s
(double mass insertion).


Before starting our analysis, a relevant remark is in order.  The
BaBar and BELLE Collaborations have recently reported the
time-dependent CP asymmetry in $B_d(\bar B_d) \to \phi K_s$. The SM
predicts such asymmetry to be the same as that measured in the $J/\psi
K_s$ decay channel. On the contrary, while $\sin 2 \beta$ as measured
in the $B \to J/\psi K_s$ channel is $0.734 \pm 0.054$ (in agreement
with the SM prediction~\cite{Stocchi:2002yi}), the combined result
from both collaborations for the corresponding $S_{\phi K}$ of $B_d
\to \phi K_s$ is $-0.39\pm 0.41$ \cite{Aubert:2002nx,Nir:2002gu} with
a $2.7 \sigma$ discrepancy between the two results. Obviously, in
particular after similar experiences in these last years, we should be
very cautious before accepting such result as a genuine indication of
NP.  Nonetheless, it is certainly legitimate to entertain the
possibility that the negative value of $S_{\phi K}$ is due to large
SUSY CP violating contributions. Then one can wonder which $\delta$'s
are relevant to produce such enhancement and, even more important,
which other significant deviations from the SM expectations in $b
\to s$ physics could be detected at $B$ factories or
hadron accelerators, if really $S_{\phi K}$ has a SUSY origin.

We remind the reader that our analysis is performed for squark and
gluino masses of $350$ GeV. We will comment at the end on what happens
if we increase such masses up to the TeV region.

\subsection{\boldmath$RR$ case}

\begin{figure}
  \begin{center}
    \begin{tabular}{c c}
      \includegraphics[width=0.48\textwidth]{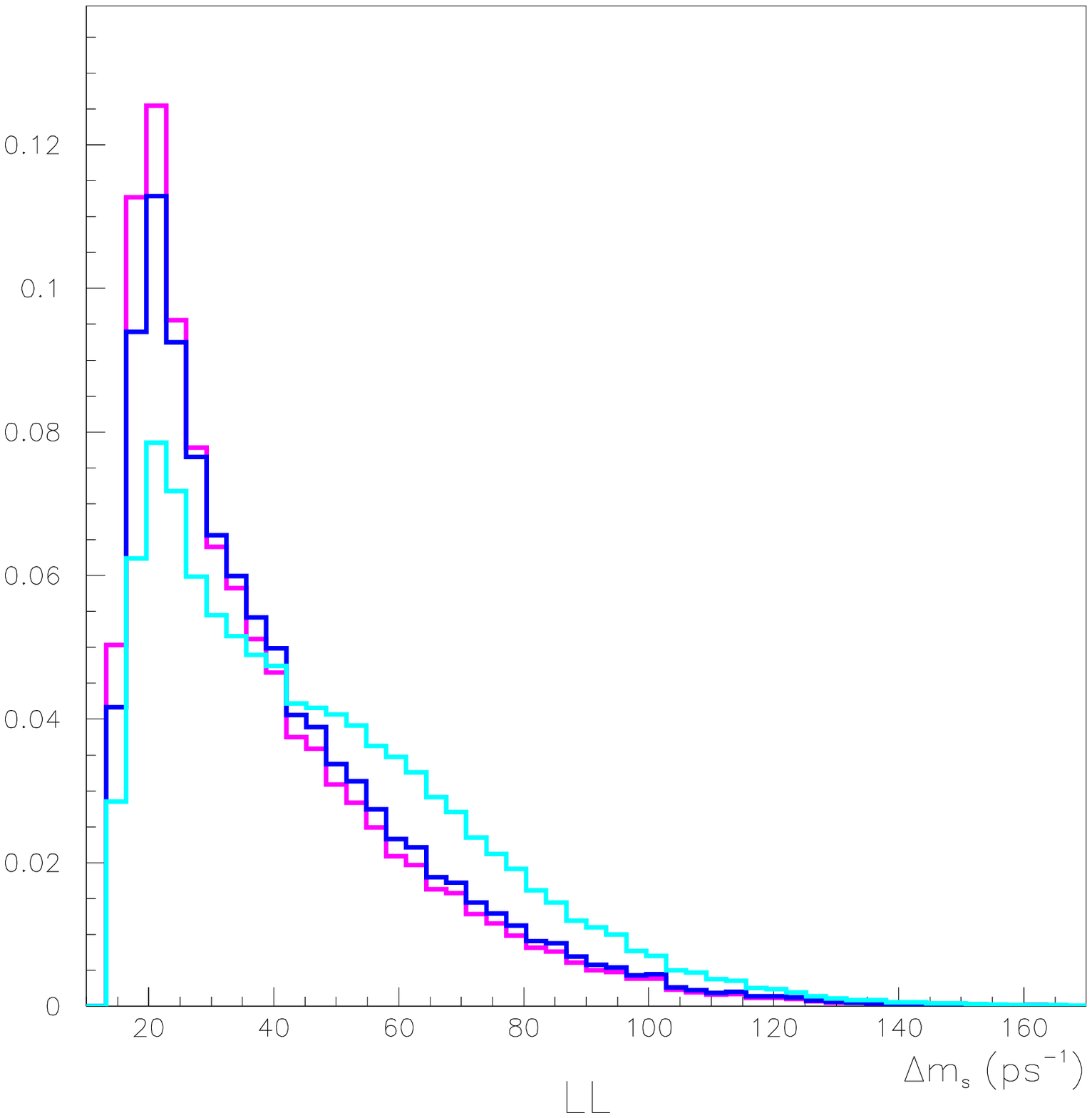} &
      \includegraphics[width=0.48\textwidth]{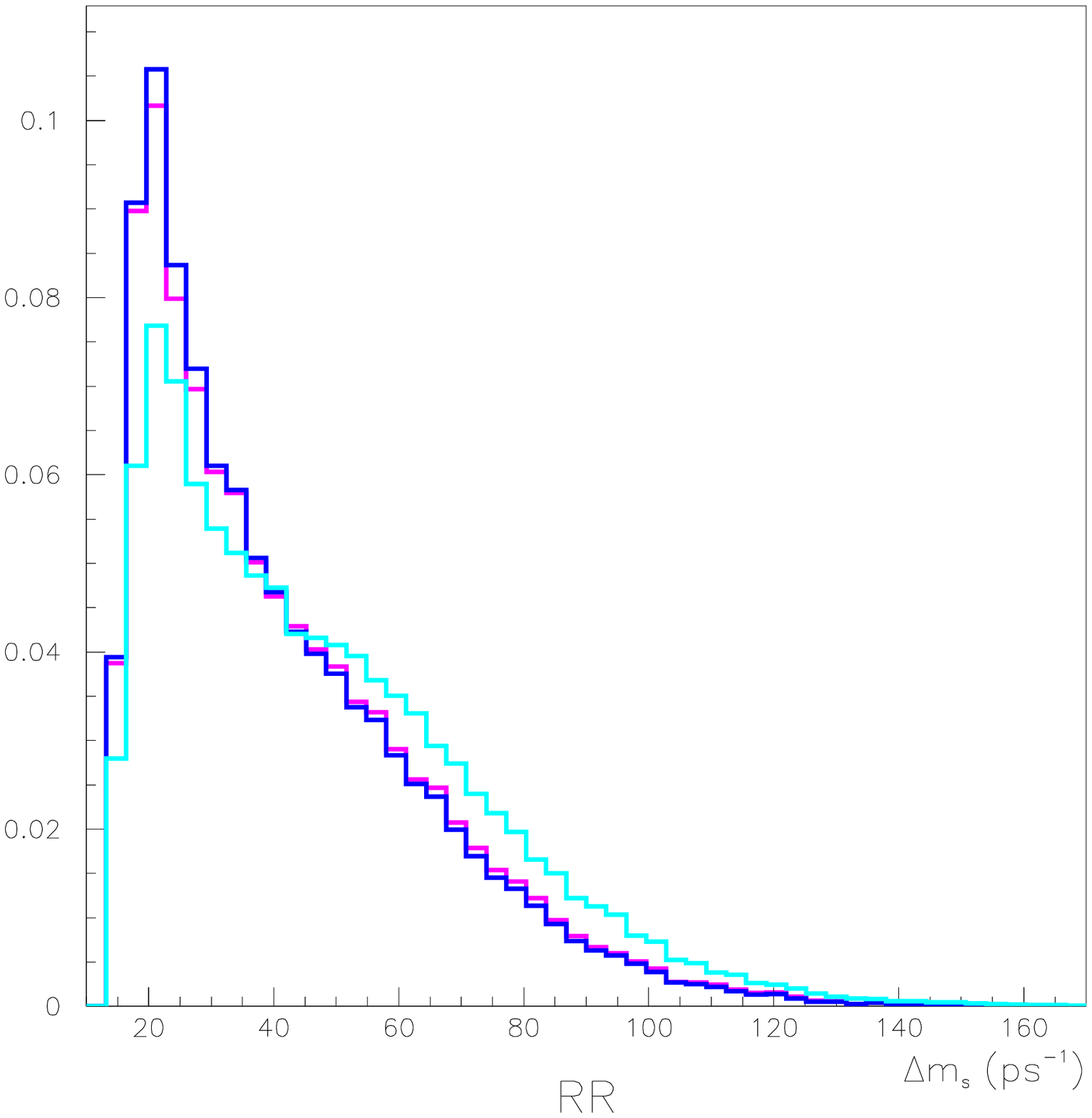} \\
      \includegraphics[width=0.48\textwidth]{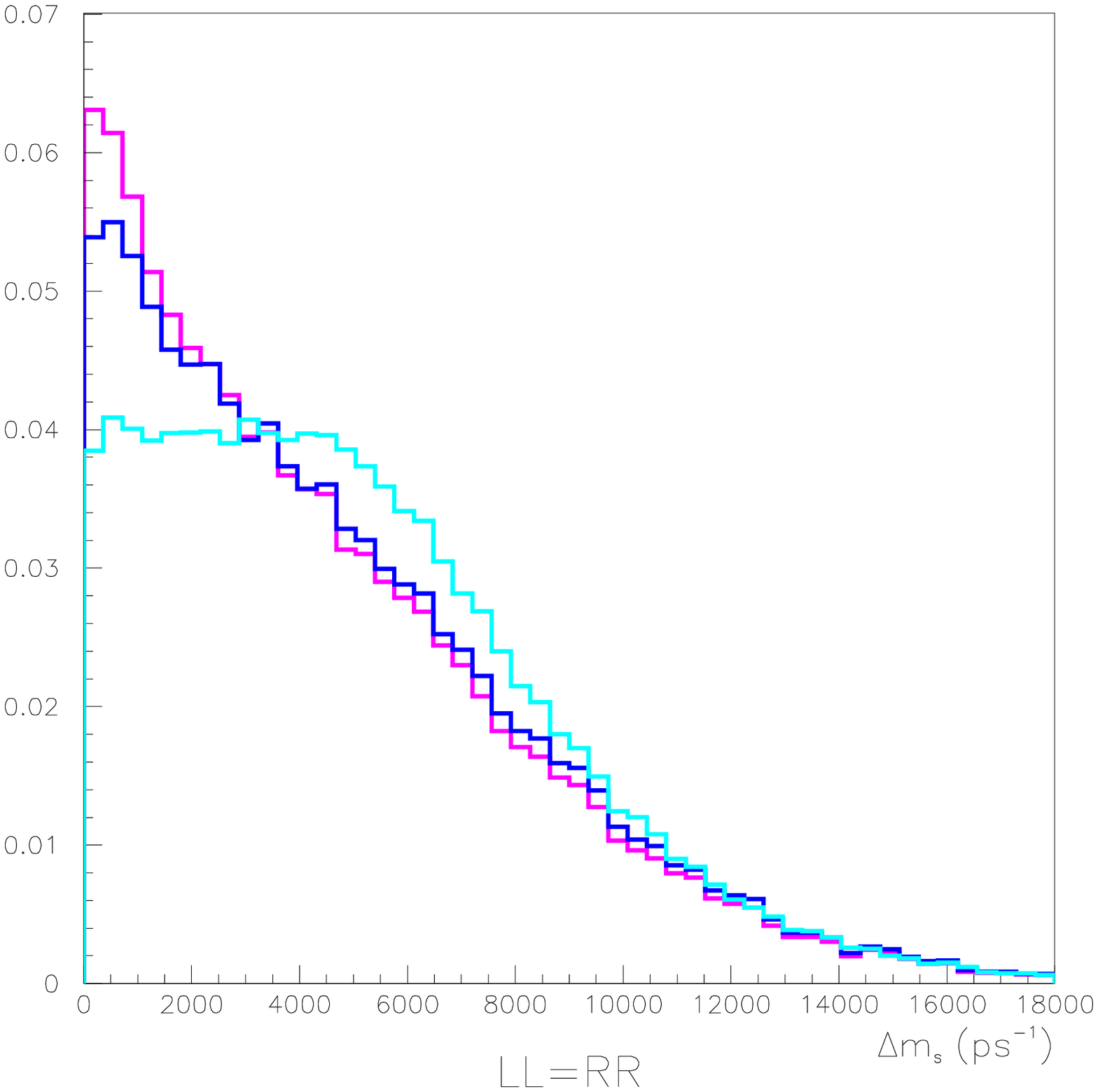} &
      \includegraphics[width=0.48\textwidth]{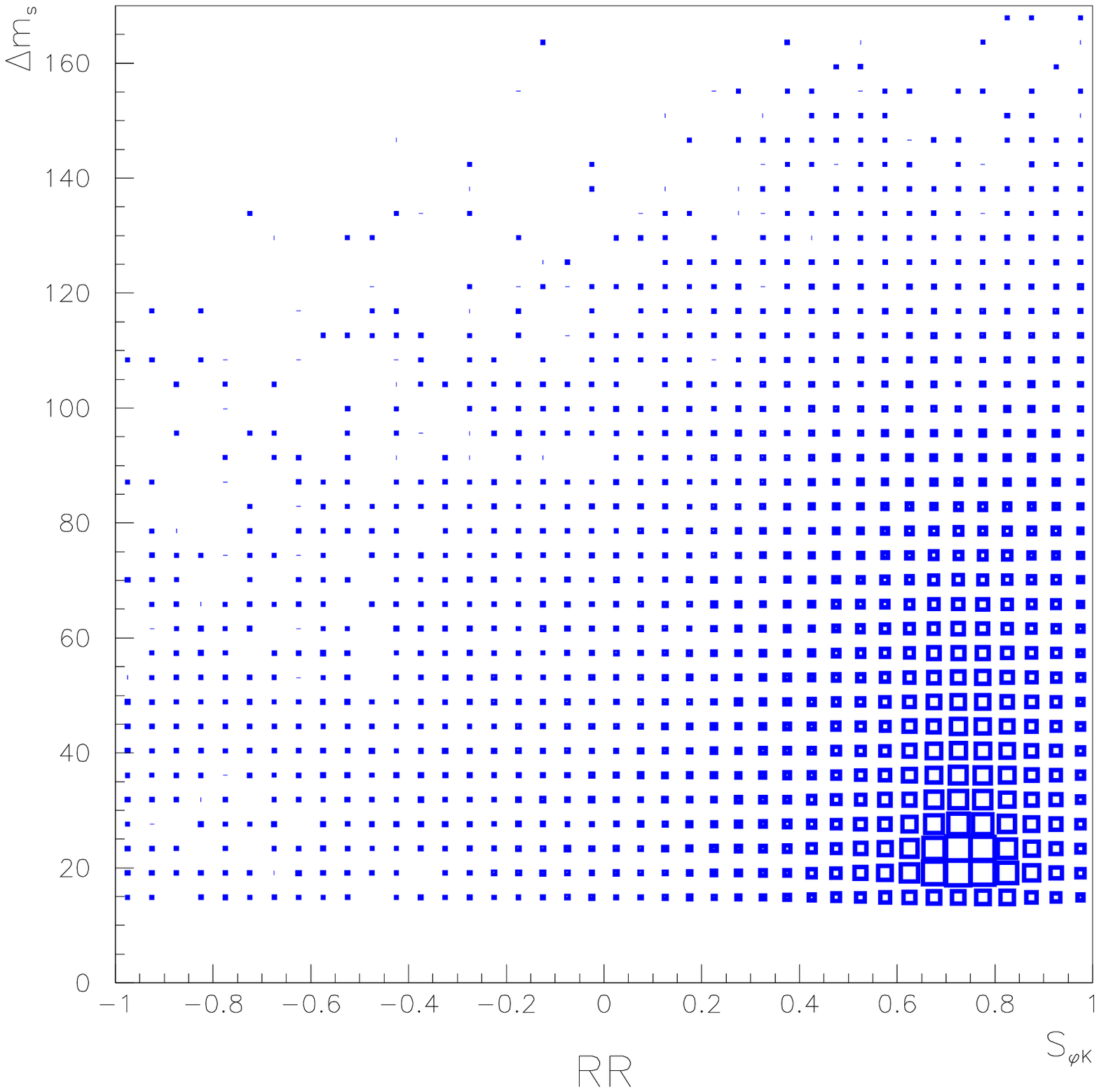} \\
      \\
    \end{tabular}
  \end{center}
  \caption{Distributions of $\Delta M_s$ for various SUSY mass 
    insertions $(\delta^d_{23})_{AB}$ with $AB=(LL,RR,LLRR)$.
    Different curves correspond to the inclusion of constraints from
    $B\to X_s\gamma$ only (magenta), $B\to X_s l^+l^-$ only (cyan) and
    all together (blue).  Lower right: correlation between $\Delta
    M_s$ and $S_{\phi K}$ in the $RR$ case.  }
  \label{fig:dms}
\end{figure}

As shown in Fig.~\ref{fig:sincos1} (upper right), although values of
$S_{\phi K}$ in the range predicted by the SM are largely favoured,
still pure $\delta_{RR}$ insertions are able to give rise to a
negative $S_{\phi K}$ in agreement with the results of BaBar and BELLE
quoted above. On this point we seem to agree with the conclusions of
ref.~\cite{Harnik:2002vs}, while being in disagreement with
refs.~\cite{Kane:2002sp} and \cite{Khalil:2002fm}. Notice that in
correspondence to negative values of $S_{\phi K}$, a vanishing $C_{
  \phi K}$, although not favoured, is however still possible. This is
due to our choice of varying the hadronic parameter $\rho_A$,
corresponding to power-suppressed annihilation contributions, over a
large range. In this way there are always particular configurations in
which the strong phase difference, and therefore the CP asymmetry,
vanish. Had we assumed $\vert \rho_A \vert < 1$ as suggested in
ref.~\cite{BBNS}, we would have found $C_{\phi K} \neq 0$ for $S_{\phi
  K}<0$~\cite{Ciuchini:2002pd}. As for the $B_s - \bar B_s$ mixing,
the distribution of $\Delta M_s$ is peaked at the SM value, but it has
a long tail at larger values, up to $\sim 120$ ps$^{-1}$ for our choice of
the range of $\delta_{RR}$. In addition, we find that the expected
correlation requiring a large $\Delta M_s$ for negative $S_{\phi K}$
is totally wiped out by the large uncertainties (see
fig.~\ref{fig:dms}, lower right). In this respect, we are at variance
with ref.~\cite{Harnik:2002vs}, where it was emphasized that if the
$RR$ squark mixing yields the large deviation from the SM for the
value of $S_{\phi K}$, then a huge contribution to the $B_s$ mixing
should necessarily follow making such oscillation unobservable at
Tevatron.  Hence, according to our analysis, in the $RR$ case it is
compatible to have a strong discrepancy between $\sin 2\beta$ and
$S_{\phi K}$ whilst $B_s-\bar B_s$ oscillations proceed as expected in
the SM (thus, being observable in the Run II of Tevatron).

As for other CP violating asymmetries to be searched for, the CP
asymmetry in $B \to X_s \gamma$ is expected to be as small as in the
SM, while, differently from the SM, we expect here in general a large
time-dependent CP asymmetry in the decay channel $B_s \to J/\psi \phi$.

\subsection{\boldmath$LL$ case}

A major difference between the $LL$ and the $RR$ cases concerns the
SUSY contributions to $B \to X_s \gamma$. When $\delta_{LL}$
dominates, SUSY contributes to the same operator which is responsible
for $B \to X_s \gamma$ in the SM and, hence, the SUSY and SM
contributions interfere. As a consequence, the rate tends to be larger
than the $RR$ case, and, moreover, a CP asymmetry can be generated up
to $5 \%$ (see fig.~\ref{fig:sinabsg}, upper left). However, given the
uncertainties, the correlation of $A_{CP}(B\to X_s\gamma)$ with
$S_{\phi K}$ is not very stringent. As can be seen from the figure,
negative values of $S_{\phi K}$ do not necessarily correspond to
non-vanishing $A_{CP}(B\to X_s\gamma)$, although typical values are
around $2\%$. Also, the constraint coming from the present measurement
of the CP asymmetry is not very effective, as can be seen for instance
from the distribution of $\Delta M_s$ in fig.~\ref{fig:dms} which is
quite similar to the $RR$ case. Finally, one expects also in this case
to observe CP violation in $B_s \to J/\psi \phi$ at hadron colliders.

\subsection{\boldmath$LR$ and \boldmath$RL$ cases}

In these cases, negative values of $S_{\phi K}$ can be easily obtained
(although a positive $S_{\phi K}$ is favoured,
cfr.~Fig.~\ref{fig:sincos1}, bottom row). The severe bound on the $LR$
mass insertion imposed by BR$(B \to X_s \gamma)$ (and $A_{CP}(B \to
X_s \gamma)$ in the $LR$ case) prevents any enhancement of the $B_s -
\bar B_s$ mixing as well as any sizeable contribution to $A_{CP}(B_s
\to J/\psi \phi)$. On the other hand, $A_{CP}(B \to X_s \gamma)$ as
large as $5-10$ \% is now attainable (Fig.~\ref{fig:sinabsg}, upper
right), offering a potentially interesting hint for NP.

Notice that the $LR$ case corresponds to a dominant $\tilde s_L - \tilde
b_R$ mass insertion, hence yielding a contribution to $b_R \to s_L
\gamma$ which interferes with the analogous leading contribution from
the SM, producing the 'hole' in fig.~\ref{fig:ranges1}, lower left.
On the contrary, the $RL$ mass insertion contributes to $b_L \to
s_R \gamma$ and thus it does not add to the leading SM amplitude.
Consequently, the CP asymmetry is as small as in the SM.

\subsection{Double mass insertion: \boldmath$(\delta_{23})_{LL}
  =(\delta_{23})_{RR}$ case}

The main feature of this case is the huge enhancement of $\Delta M_s$
which is made possible by the contribution of the double insertion $LL$
and $RR$ in the box diagrams to operators with mixed chiralities
(Fig.~\ref{fig:dms}, lower left). Differently from all the previous
cases, we are facing a situation here where $A_{CP}(B \to \phi K_s)$ at
its present experimental value should be accounted for by the presence
of SUSY, while $\Delta M_s$ could be so large that the $B_s - \bar
B_s$ mixing could escape detection not only at Tevatron, but even at
BTeV or LHCB. Hence, this would be a case for remarkable signatures of
SUSY in $b \to s$ physics.

\subsection{Double mass insertion: \boldmath$(\delta_{23})_{RL}
  =(\delta_{23})_{RR}$ case}

As we previously mentioned, in MSSM realizations obtained as
low-energy limits of SUSY GUT's where neutrinos and right-handed down
quarks live in the same SU(5) fiveplets, the case for a large
$(\delta_{23}^d)_{RR}$ is strongly motivated by the observed large
mixing between neutrinos of the second and third generation.  In
addition, we can have also a large $\tilde b_R - \tilde b_L$ if the
product $\mu \tan \beta$ happens to be enhanced (by large values of
$\mu$ and/or $\tan \beta$)\footnote{This is however the case in which
  chargino-mediated contributions are expected to be important and
  should be included choosing a specific model. Therefore, in this
  case the results of our model-independent analysis should be
  interpreted with care.}.  If this is the case, the subsequent $\tilde
b_L - \tilde b_R$ and $\tilde b_R - \tilde s_R$ mass insertions lead
to a large $\tilde b_L - \tilde s_R$ transition, i.e.  a large
$(\delta_{23}^d)_{LR}$. This makes of interest to consider the case
where $(\delta_{23}^d)_{RR}$ and $(\delta_{23}^d)_{RL}$ are both
simultaneously large.  Phenomenologically, there is no big difference
between this case and the $RL$ case, as can be seen from
figs.~\ref{fig:ranges2}--\ref{fig:sinim2}.

We close this section by remarking that in the $LR$, $RL$ and $RL=RR$
cases, since for $m_{\tilde g}=m_{\tilde q}=350$ GeV the constraints
on the $\delta_{23}^d$'s are of order $1 \times 10^{-2}$, it is obvious
that the same phenomenology in $\Delta B=1$ processes can be obtained
at larger values of mass insertions and of squark and gluino masses.
In this case, contributions to $\Delta B=2$ processes become more
important for larger masses. One could naively think of going up to
$3.5$ TeV with $\delta^d_{23} \sim 1$, however so large values of
$\delta_{23}^d$ and $m_{\tilde q}$ produce charge and color breaking
minima~\cite{Casas:1996de}. In the remaining cases, where the limits
on $\delta_{23}^d$ at $m_{\tilde g}=m_{\tilde q}=350$ GeV are of order
$1$, the SUSY effects clearly weaken when going to higher values of
sparticle masses. 

\section{Outlook}
 
All the known FCNC and CP violation phenomenology in kaon, beauty,
charm, lepton physics and the EDM's confirm the simple CKM picture
of flavor physics as provided by the SM.

This may be interpreted as an indication that the relevant new physics
involved in the electroweak symmetry breaking is flavor blind or that
the new particles associated with it are actually relatively heavy,
say at least beyond the TeV threshold.

However, sticking more strictly and cautiously to what has been
experimentally ascertained so far, one should actually conclude that
although FCNC physics involving the first two generations (i.e.~$s
\to d$ transitions) can hardly offer any prospect of
``visibility'' for NP, the same cannot be said for FCNC and
CP violation involving the third generation, in particular as far as
$b \to s$ transitions are concerned. The first result on
$A_{CP}(B \to \phi K_s)$ from BaBar and Belle is an example that some
room for surprise is still available.

In this paper we analyzed the prospects for the $b \to s$
physics for the particularly interesting case where the low-energy new
physics is represented by a generic MSSM. Minimality refers here only
to the minimal amount of superfields needed to supersymmetrize the SM
and to the presence of R parity. Otherwise the soft breaking terms are
left completely free and constrained only by phenomenology.
Technically the best way we have to account for the SUSY FCNC
contributions in such a general framework is via the mass insertion
method using the leading gluino exchange contributions.

Our results in a generic MSSM confirm that FCNC and CP violation in
physics involving $b \to s$ transitions still offer
opportunities to disentangle effects genuinely due to NP. In
particular the discrepancy between the amounts of CP violation in the
two $B_d$ decay channels $J/\psi K_s$ and $\phi K_s$ can be accounted
for in the MSSM while respecting all the existing constraints in $B$
physics, first of all the $BR(B \to X_s \gamma)$. The relevant question
is then which processes offer the best chances to provide other hints
of the presence of low-energy SUSY.

First, needless to say, it is mandatory to further assess the
time-dependent CP asymmetry in the decay channel $B \to \phi K_s$.
Should the abovementioned discrepancy signaling NP be firmly
confirmed, then this process would become decisive in discriminating
among different MSSM realizations. Although, as we have seen, it is
possible to reproduce the negative $S_{\phi K}$ in a variety of different
options for the SUSY soft breaking down squark masses, the allowed
regions in the SUSY parameter space to obtain this result are more or
less tightly constrained according to the kind of $\delta_{23}^d$ mass
insertion which is dominant.

We think that in order of importance after the reassessment of $A_{CP}
(B \to \phi K_s)$, comes the measurement of the $B_s - \bar B_s$
mixing. Finding $\Delta M_s$ larger than $20$ ps$^{-1}$ would hint at
NP and, in our context, would imply that the chirality-changing mass
insertions or the $RL=RR$ double insertion should not be dominant in
$b \to s$ transitions. $RR$ or $LL$ could account for a $\Delta M_s$
up to $\sim 120$ ps$^{-1}$. Larger values would call for the double
insertion $LL=RR$ to ensure such a huge enhancement of $\Delta M_s$
while respecting the constraint on $BR(B \to X_s \gamma)$. An
interesting, alternative prospect would arise in case $\Delta M_s$ is
found as expected in the SM while, at the same time, $S_{\phi K}$ is
confirmed to be negative. This scenario would favour the $RL$ or $LR$ 
possibility, even though all other cases but $LL=RR$ do not
necessarily lead to large $\Delta M_s$.

Keeping to $B$ physics to be studied at $B$ factories, we point out
that the CP asymmetry in $B \to X_s \gamma$ remains of utmost
interest.  As we know, this asymmetry is so small in the SM that it
should not be possible to detect it. We have seen that in particular
with $LR$ $b \to s$ insertions such asymmetry can be enhanced up to
$10$ \% making it possibly detectable in a not too distant future.

Finally, once we will have at disposal large amounts of $B_s$ at
hadron colliders it will be of great interest to study processes which
are expected to be mostly CP conserving in the SM, while they are
expected to receive possibly large contributions from SUSY in the $b
\to s$ transitions. Indeed, in the SM the amplitude for $B_s - \bar
B_s$ mixing is dominated by the top quark exchange and, hence, it does
not have an imaginary part up to doubly Cabibbo suppressed terms.
Also the transition $b \to s$ through top and $W$ exchange does not
exhibit the CKM phase and, hence, no leading CP contribution should
arise in amplitudes for $b \to s$ processes. In conclusion decays like
$B_s \to J/\psi \phi$ should have a negligible amount of CP violation.
Quite on the contrary, if the measured negative $S_{\phi K}$ is due to
a large, complex $\delta_{23}^d$ mass insertion, we expect some of the
above processes to exhibit a significant amount of CP violation. In
particular in the case of $RR$ insertions both the $b \to s$
amplitudes and the $B_s$ mixing would receive non negligible
contributions from Im$\,\delta_{23}^d$, while in the case where $S_{
  \phi K}$ is dominated by $RL$ or $LR$ insertions we do not expect
any sizable contribution to $B_s$ mixing. Still, we can expect a
potentially interesting contribution to CP violation in the $B_s \to
J/\psi \phi$ decay amplitude.
  
We hope that we made the case for $b \to s$ physics in the search for
indirect SUSY signals quite clear. Still someone could object that at
least some of the relevant probes of $b \to s$ transitions in $B_s$
physics are likely to have to wait for the advent of BTeV or LHC to
become feasible. And at that point, one would be tempted to say, we
will know whether low energy SUSY is there just by direct tests
without invoking the indirect searches that we advocated in this work.
We do not agree with this position: even if SUSY particles are
detected, we are convinced that FCNC and CP violation, in $b \to s$
physics in particular, will constitute an ideal ground to have a
further access to the SUSY structure and will be complementary to the
other direct forms of SUSY studies.

\section*{Acknowledgments}
We are much indebted to G.~Martinelli for his participation to the
early stage of this work. We thank P.~Gambino for fruitful
discussions.  We also thank D.~Larson for providing us with helpful
details on the analysis of ref.~\cite{Harnik:2002vs}.

\end{document}